\documentclass [11pt]{article}
\usepackage{amsmath,amsthm,amsfonts,amscd,eucal,latexsym,amssymb}
\usepackage{epsfig}  
\usepackage[latin1]{inputenc}
\usepackage{verbatim} 
\def\cA{{\ca A}}

\def\cC{{\ca C}}
\def\cD{{\ca D}}
\def\cE{{\ca E}}
\def\cF{{\ca F}}

\def\cH{{\ca H}}
\def\cI{{\ca I}}

\def\cO{{\ca O}}

\def\mA{{\mathcal A}}
\def\mD{{\mathcal D}}

\def\mF{{\mathcal F}}

\def\mO{{\mathcal O}}

\def\bC{{\mathbb C}}           

\def\bK{{\mathbb K}}

\def\bA{{\mathbb A}}
\def\bN{{\mathbb N}}

\def\bR{{\mathbb R}}

\def\bZ{{\mathbb Z}}

\def\gF{{\mathfrak F}}

\def\gH{{\mathfrak H}}

\def\beq{\begin{eqnarray}}
\def\eeq{\end{eqnarray}}

\def\at{\left(}               

\def\ct{\right)}              
\newcommand{\ca}[1]{{\cal #1}}         

\def\alp{\alpha}
\def\be{\beta}
\def\ga{\gamma}
\def\de{\delta}
\def\ep{\varepsilon}

\def\la{\lambda}
\def\ro{\varrho}
\def\si{\sigma}
\def\om{\omega}
\def\ph{\varphi}

\def\Ga{\Gamma}
\def\De{\Delta}
\def\La{\Lambda}
\def\Si{\Sigma}
\def\Om{\Omega}

\def\Th{\Theta}

\def\dnot{{\displaystyle{\not}}}

\newcommand{\nref}[1]{(\ref{#1})}

\newcommand{\abs}[1]{\left|{#1}\right|}
\newcommand{\wick}[1]{:\!{#1}\!:}

\newcounter{proposition}[section]
\newcounter{theorem}[section]
\newcounter{lemma}[section]
\newcounter{definition}[section]
\newcounter{corollary}[section]
\def\theproposition{\thesection.\arabic{proposition}}
\def\thetheorem{\thesection.\arabic{theorem}}
\def\thelemma{\thesection.\arabic{lemma}}
\def\thedefinition{\thesection.\arabic{definition}}
\def\thecorollary{\thesection.\arabic{corollary}}

\newcommand{\se}[1]{\section{#1}}

\def\vsp{\vspace{0.2cm}}
\def\vspp{\vspace{0.1cm}}

\def\sse #1 {\vsp\ifhmode{\par}\fi\refstepcounter{subsection}
  \noindent {\bf\thesubsection}. {\em #1}.\quad
  \addcontentsline{toc}{subsection}{\protect\numberline{\thesubsection} #1}%
  }

\def\ssb #1 {\vsp\ifhmode{\par}\fi\refstepcounter{subsection}
  \noindent {\bf\thesubsection.} {\bf #1.}\quad
  \addcontentsline{toc}{subsection}{\protect\numberline{\thesubsection} #1}%
  }

\def\ssa #1 {\ifhmode{\par}\fi\refstepcounter{subsection}
  \noindent {\bf\thesubsection.} {\bf #1.}\quad
  \addcontentsline{toc}{subsection}{\protect\numberline{\thesubsection} #1}%
  }

\def\proposizione #1 {\vsp\vspp\ifhmode{\par}\fi\refstepcounter{proposition}
  \vsp\ifhmode{\par}\fi\noindent {\bf Proposition \theproposition}. \quad {\em #1}\vsp\vspp}
\def\teorema #1 {\vsp\vspp\ifhmode{\par}\fi\refstepcounter{theorem}
  \vsp\ifhmode{\par}\fi\noindent {\bf Theorem \thetheorem}. \quad {\em #1}\vsp\vspp}
\def\lemma #1 {\vsp\vspp\ifhmode{\par}\fi\refstepcounter{lemma}
  \vsp\ifhmode{\par}\fi\noindent {\bf Lemma \thelemma}. \quad {\em #1}\vsp\vspp}
\def\definizione #1 {\vsp\vspp\ifhmode{\par}\fi\refstepcounter{definition}
  \vsp\ifhmode{\par}\fi\noindent {\bf Definition \thedefinition}. \quad {\em #1}\vsp\vspp}
\def\corollario #1 {\vsp\vspp\ifhmode{\par}\fi\refstepcounter{corollary}
  \vsp\ifhmode{\par}\fi\noindent {\bf Corollary \thecorollary}. \quad {\em #1}\vsp\vspp}

\def\proof #1 {\vsp\ifhmode{\par}\fi\noindent {\it Proof.} {#1} $\Box$\vsp\par}
\def\remark #1 {\vsp\vspp\ifhmode{\par}\fi\noindent\noindent {\bf Remark.} {#1}\vsp\vspp\par}
\def\remarks #1 {\vsp\vspp\ifhmode{\par}\fi\noindent\noindent {\bf Remarks.} {#1}\vsp\vspp\par}

\begin{document}

\hfill{\sl Desy 09-049, ZMP-HH/09-9 - March 2009}
\par
\bigskip
\par
\rm


\par
\bigskip
\LARGE
\noindent
{\bf The extended algebra of observables for Dirac fields and the trace anomaly of their stress-energy tensor}
\bigskip
\par
\rm
\normalsize


\large
\noindent {\bf Claudio Dappiaggi$^{1,a}$}, {\bf Thomas-Paul Hack$^{1,b}$}
and {\bf Nicola Pinamonti$^{1,c}$}\\
\par
\small
\noindent $^1$
II. Institut f\"ur Theoretische Physik, Universit\"at Hamburg,
Luruper Chaussee 149,
D-22761 Hamburg, Germany.\smallskip

\noindent $^a$  claudio.dappiaggi@desy.de, $^b$ thomas-paul.hack@desy.de, $^c$  nicola.pinamonti@desy.de\\
 \normalsize

\par

\rm\normalsize
\noindent {\small Version of \today}

\rm\normalsize


\par
\bigskip

\noindent
\small
{\bf Abstract}.
We discuss from scratch the classical structure of Dirac spinors on an arbitrary globally hyperbolic, 
Lorentzian spacetime, their formulation as a locally covariant quantum field theory, and the associated notion of a Hadamard state. 
Eventually, we develop the notion of Wick polynomials for spinor fields, and we employ the latter to construct a covariantly conserved stress-energy tensor suited for back-reaction computations. We shall explicitly calculate its trace anomaly in particular.
\normalsize

\vskip .3cm

\noindent{\em Dedicated to the memory of our friend and colleague Raffaele Punzi}

\tableofcontents

\se{Introduction}

In the last few decades we witnessed an amazing leap forward in our understanding of the formulation of 
quantum field theory on curved backgrounds thanks to the efforts of many research groups which often 
tackled this topic by means of the algebraic formalism. A careful analysis of all the related achievements, 
though tempting, would require a review on its own and, instead, we shall restrict ourselves to briefly 
mentioning the results of a recent manuscript, which has prompted our interest towards the topic discussed in 
this paper. To wit, in \cite{dfp}, it was shown that, in the framework of semiclassical Einstein's equations,
it is possible to construct explicit solutions for a homogeneous and isotropic Friedmann-Robertson-Walker spacetime with flat spatial sections,  where the assumption has been made that the matter content is described by a suitably quantised free massive scalar field. In order to prove this result, two ingredients have played a key role, namely, Hadamard states, as the natural candidates for a ground state, and the quantum behaviour of the regularised stress-energy tensor $T_{\mu\nu}$. Particularly, as end point of the above cited paper, a late times stable de Sitter solution has been displayed and, hence, an effective cosmological constant has arisen without inserting it from the very beginning as an input datum. If one tries to seek the origin of this genuine quantum effect, one can realise that it is ultimately rooted in the so-called {\it trace anomaly}, 
{\it i.e.}, in a few words, the expectation value of the regularised trace is not vanishing on Hadamard states, even though this is the case at a classical level. 

Although interesting, the derivation of the aforementioned result rises naturally the question about its robustness since one could wonder if this behaviour is a feature pertaining only to scalar fields or if it holds true for any kind of matter constituent. It thus seems advisable to try to apply the same scheme of reasoning in the context of a free Dirac field. To our utmost surprise, we have realised that the accomplishment of this goal has not been as easy as one might {\it a priori} believe since, just by a quick scan of the available literature, it is manifest how spinor fields play a somehow ancillary role in the arena of algebraic quantum field theory.
In fact, there are only few mathematically sound results and many tools and concepts, which have been thoroughly discussed for scalar fields, have barely been scratched for spinors. As an example, let us recall that we are mostly  concerned with the trace anomaly for Dirac fields and it turns out that this quantity has indeed been already investigated in the late seventies in \cite{Christensen, chris1}, though by means of the so-called {\it DeWitt-Schwinger expansion} which lacks mathematical rigour and, therefore, we cannot call our understanding of this anomaly complete yet. 

With this in mind, before we can tackle any specific topic in a cosmological scenario, our first concern must lie in the amendment of the the above mentioned problem. This will indeed be the main point of the paper and we shall discuss it within the framework of the algebraic approach to quantum field theory. To this end, we must also take into account that the scientific community interested in this problem might not be acquainted with the formulation of a spinor field theory on a curved background, which, already at a classical level, turns out to be rather different and more complicated than the scalar counterpart. Hence, as a starting point, we shall review the construction of a Dirac field in a classical framework emphasising the role of the underlying geometric structures which are needed in order to fully describe both the kinematically and the dynamically allowed field configurations. To this effect, our analysis will benefit from earlier works which have already dwelled on this topic and, most notably, we shall refer to the seminal paper of Lichnerowicz \cite{Lichnerowicz} as well as to \cite{Dimock, Sanders}.

Subsequently, we shall discuss the quantisation of a Dirac field on a curved background and, in this respect,
one should mention that there are several possibilities at our disposal. On the one hand one could follow the
point of view already suggested in \cite{Dimock}, while, on the other hand, one could also analyse the
problem from the perspective of Araki \cite{Araki70}, whose scheme has the peculiarity of unifying spinors
and cospinors in a single body before quantising them. This leads to a natural definition both of a CAR $*$-algebra of fields and of a subalgebra of observables, once we require at least that  elements whose supports are spacelike separated must commute. Furthermore, this scheme, also at the heart of \cite{Verch, Koehler, Kratzert, Hollands, Fewster, Sanders}, has the remarkable advantage of being well suited to recast the quantum theory of Dirac fields in the language of {\it Locally Covariant Quantum Field Theory} \cite{BFV}, as we will point out. 

In order to fully control the machinery of a quantum field, the scalar field scenario already thought us that
one has to understand well which algebraic states one should use and, to this avail, the ones of Hadamard
type are the natural choice in the context of Dirac fields, too; these play the role of ground states in a
curved background and their ultraviolet behaviour closely mimics that of the Minkowski spacetime vacuum
state. Consequently, the fluctuations of the components of the stress-energy tensor on these states are
bounded, a property which is vital in the context of semiclassical Einstein's equations. Hadamard states for Dirac fields have already been discussed in \cite{Koehler, Verch, Hollands, Kratzert, SahlmannVerch, DaHo} and we will review their properties in detail before employing them to achieve the first of our main results, namely, the construction of the extended algebra of Wick polynomials in a spinor framework. To this end, we will follow the path paved in the scalar scenario in \cite{BFK, BF00, HW01, HW02} where it has been displayed that such polynomials lie at the basis of a sound S-matrix formalism for interacting field theories on a globally hyperbolic curved background. 

Notwithstanding, Wick polynomials are already valuable in free field theories and we will, as already
anticipated, indeed use them to  achieve our second main result, {\it i.e.}, the definition of a well-behaved quantum stress-energy tensor operator $T_{\mu\nu}$ for Dirac fields. To achieve this goal, we shall follow a procedure similar to that discussed in \cite{Mo03} (see also the related work \cite{HW04}) for a free scalar field, {\it i.e.}, we shall introduce an improved point-splitting procedure
to define $T_{\mu\nu}$ evaluated on a Hadamard state. This leads to a new overall stress-energy tensor which 
does not alter classical dynamics and is ultimately conserved at the quantum level. Furthermore, as a 
by-product of our analysis, we shall also be able to explicitly compute the expectation value of its trace 
which will agree, up to terms proportional to $\square R$, with previously found results while being derived 
in a rigorous framework.

\se{Dirac fields: a classical overview}\label{classical}
Since, as we have outlined in the introduction, the aim of this paper is to provide an as much as possible self-contained
approach to some topics related to the quantum description of Dirac fields in curved backgrounds, we will start with a description of Dirac spinors in a classical framework. Although such topic has been already discussed both from a geometrical and from an analytical point of view by many authors, we reckon the we should try to recall the main features of the 
classical approach in order both to facilitate the understanding of the quantum aspects and to fix some subtleties which ubiquitously arise in these scenarios.

\subsection{On the spin structure and related geometric entities}
Bearing in mind this overall philosophy, we shall mainly devote this subsection to the introduction of spin
structures and of the Dirac bundles, in order to characterise (co)spinors as suitable sections. We shall not
dwell too much on the geometrical contents and for the potential readers who might find our approach too
shallow we present our apologises and point them to \cite{Lawson} for a careful discussion of most of the
forthcoming concepts and applications. 

As a starting point, let us fix that, in this paper, a spacetime is meant to be a
four-dimensional, Hausdorff, smooth manifold endowed with a Lorentzian metric, whose
signature is chosen as $(-,+,+,+)$. Furthermore, since it is common wisdom to associate Dirac fields
to the notion of spin, we need a few definitions as a first step:

\definizione\label{sping}{We call {\bf spin group} $Spin(p,q)$ with $p,q\in\mathbb{N}$ the double cover of
$SO(p,q)$, {\it i.e.}, it exists the following short exact sequence:
$$\left\{e\right\}\longrightarrow\mathbb{Z}_2\longrightarrow Spin(p,q)\longrightarrow SO(p,q)\longrightarrow
\left\{e\right\},$$
where $\left\{e\right\}$ stands for the trivial group, whereas $\mathbb{Z}_2\doteq\left\{\pm 1\right\}$ is the
cyclic group of order $2$. Therefore, any element of Spin(p,q) induces an element of SO(p,q). Such a surjective
covering will be indicated as $\Pi:Spin(p,q)\to SO(p,q)$.}

\remark{As a consequence of the above definition, it exists an isomorphism between $Spin(p,q)$ and 
$Spin(q,p)$ for any possible value of both $p$ and $q$. Furthermore, it is also possible to talk about the dimension of such classical Lie group which, per direct inspection, is 
$$dim\left(Spin(p,q)\right)=\frac{1}{2}(p+q)(p+q-1).$$
Furthermore, for all $p,q>0$, the spin group has two connected components, where we denote the component connected to the identity as $Spin_0(q,p)$. The latter insight entails that the scenario
with $p=3$ and $q=1$ is of great interest since, in this case, $Spin_0(1,3)$ is isomorphic to $SL(2,\mathbb{C})$. }

The above definition represented only the first step toward the definition of a Dirac field since, in modern
classical field theory, the geometric interpretation of a kinematically allowed configuration is that of a 
section of a suitable associated bundle. Within this respect, one should notice that, (un)fortunately, in the 
often analysed case of a scalar field, such nice perspective does not really enter the fray, whereas, in this
case, such a luxury is not at our disposal, being a spinor intrinsically a vector-valued field. Hence, we shall
now show how the notion of spin group can be intertwined with that of a classical field. As a first step we
need further definitions.

\definizione{Given a vector bundle $E\doteq E[V,\pi,M]$ over a Lorentzian manifold M with typical fibre $V$
and projection map $\pi:E\to M$, we call {\bf frame} $F(x)$ over the point $x$ the assignment of an
ordered basis to the fibre $\pi^{-1}(x)\equiv V$, {\it i.e.}, a map $p:\mathbb{K}^k\to\pi^{-1}(x)$, being $k$
the dimension of $V$ and $\bK\in\{\bR,\bC\}$. Furthermore, we require that the right action of the group $GL(k,\mathbb{K})$ on the
fibre is both free and transitive.}

\remark{Such a definition is at the heart of the widely exploited concept of tetrads in general relativity
and, on a practical ground, it is remarkable since it guarantees us that whenever a vector bundle is
associated to the underlying spacetime one is free to choose a basis of such a space and that the
rules of changing the basis are left untouched; this is encoded in the action of $GL(k,\mathbb{K})$, {\it i.e.},
each element of this group transforms a basis into another one. Furthermore, the request of transitivity
guarantees us that it is always possible to transform any chosen basis into any other one, whereas demanding 
a free action entails that only the identity element leaves a chosen basis unchanged. The most 
notable and ubiquitously used application of such a definition is the \emph{tangent bundle} where 
$V\equiv\bR^k$, being $k=dim M$. In the following, we shall have this case in mind and hence we shall identify
$E$ as $E\equiv TM=TM[\bR^k,\pi,M]$.}

Therefore, if we call $F_xM\sim GL(k,\mathbb{R})$ the set of all
possible frames over a point $x\in M$, we can gather all this information into a unique object:

\definizione\label{framebun}{A {\bf frame bundle} $FM$ associated to $TM$ is $FM\doteq \bigsqcup_xF_xM$. This
is a fibre bundle on its own, namely $FM=FM[GL(k,\bR), \pi',  M]$, with $GL(k,\bR)$ as typical fibre and the
projection map $\pi':FM\to M$. Whenever the base space $M$ is an orientable and time orientable spacetime, hence also endowed 
with a (local) \underline{non-degenerate} metric of signature $(p,q)$, we can choose a space and time orientation, and hence reduce $GL(k,R)$ to $SO_0(p,q)$ with $p+q=k$. We shall always consider such case in the following.}

\remark{A fibre bundle as above is an example of a \emph{principle fibre bundle}, having the same group as both the typical fibre and the structure group. By requiring the right action of the group to leave the base point invariant, {\it e.g.}, $\pi' \circ R_{GL(k,\bR)}=\pi'$ for $FM$, one can straightforwardly extend it to the full principle fibre bundle. This extension is independent of local trivialisations, since these are related by the left action of the group and the left and right action of a group on itself commute. One can even switch perspectives and define the fibres and base points of a principle fibre bundle via the orbits of the right action.}

We are now in the position to eventually introduce the main geometric structure of this paper which
lies at the heart of the construction and of the analysis of a Dirac (co)spinor:

\definizione\label{spinstr}{Given a manifold $M$ with a non degenerate metric of signature $(p,q)$, a 
{\bf spin structure} is the pair $(SM,\rho)$ where $SM\doteq SM[Spin_0(p,q),\widetilde\pi,M]$ is a principle  fibre bundle 
over $M$ with the identity component of the spin group as typical fibre. Moreover, $\rho$ is a bundle homomorphism from $SM$ to $FM$
subject to the following conditions:
\begin{enumerate}
\item $\rho$ is base point preserving, such that
$$\pi'\circ\rho =\widetilde\pi,$$
\item $\rho$ must be equivariant, i.e., calling $R_{\widetilde\Lambda}$ and $R_\Lambda$ the natural right
actions of $Spin_0(p,q)$ on $SM$ and of $SO_0(p,q)$ on $FM$ respectively, we require that  
$$\rho\circ R_{\widetilde\Lambda}=R_{\Lambda}\circ\rho,\qquad\forall\Lambda\in SO_0(p,q),$$
where $\Lambda=\Pi(\widetilde\Lambda)$, being $\Pi$ as in definition \ref{sping}.
\end{enumerate}}

\remark{A natural, apparently na\"ive, but ultimately rather complicated question a potential reader might 
ask, is why one should cope with such complicated geometric structures to deal with a somehow natural concept 
such as that of spin. To answer this question, let us keep in mind the notion of a Dirac field in the flat scenario and 
try to generalise it to a curved background with Occam's razor as a principle. That said, in four-dimensional Minkowski spacetime, one can employ a standard construction, dating from the late thirties and 
due to Wigner, according to which a spin $\frac{1}{2}$ field is nothing but a suitable function satisfying 
the Dirac equation and transforming under a unitary and parity invariant representation of the universal
cover of the Poincar\'e group induced from  the $j=\frac{1}{2}$ representation of $SU(2)$ (see chapter 21 of
\cite{Barut} for more details). It is remarkable to notice that, once restricted to $SL(2,\mathbb{C})$, {\it 
i.e.}, neglecting the translations, this representation coincides with the $D^{\left(\frac{1}{2},0\right)}
\oplus D^{\left(0,\frac{1}{2}\right)}$ representation of the aforementioned group. Although there is no 
translational invariance in a curved background, the previous definitions grants us that, by means of the 
spin structure, it is nonetheless possible to define, barring certain technical requirements, a natural 
notion of an $SL(2,\bC)$-group associated to a spacetime. Furthermore, the very absence of translational invariance implies that such a definition of spin in curved backgrounds seems really the best one can do.}

Hence, though not sufficient to determine full dynamics of a Dirac spinor, our philosophy will be to seek
to characterise the kinematically allowed configurations of such a field by means of the mentioned $D^{\left(
\frac{1}{2},0\right)}\oplus D^{\left(0,\frac{1}{2}\right)}$ representation, while remembering at the same 
time that, classically, fields should be understood as sections of a suitable vector bundle. To combine these two concepts, we proceed in the following way:

\definizione{We call {\bf Dirac bundle} of a four dimensional Lorentzian spacetime M with respect to the
representation $T\doteq D^{\left(\frac{1}{2},0\right)}\oplus D^{\left(0,\frac{1}{2}\right)} $ of $SL(2,\bC)$ 
on $\bC^4$ the associated bundle $DM\doteq SM\times_T\bC^4$. This is the set of equivalence classes $[(p,
z)]$, where $p\in SM$, $z\in\bC^4$ and equivalence is defined out of the relation
$$(p_1,z_1)\sim(p_2,z_2),$$
if and only if it exists an element $A$ of $SL(2,\bC)$ such that $L_A(p_1)=p_2$ and $T(A^{-1})z_1=z_2$, where 
$L_A$ denotes the left action of $A$ on $SM$. The global structure of
$DM$ is that of a fibre bundle over $M$ with typical fibre $\bC^4$, and the projection map $\pi_D$ is traded
from the one of $SM$, namely, $\forall\,[(p,z)]\in DM$, it holds
$$\pi_D[(p,z^i)]\doteq\widetilde\pi(p).$$ Furthermore, if we endow $\bC^4$ with the standard non degenerate internal product, we can 
construct the {\bf dual Dirac bundle} $D^*M$ as the $\bC^{4*}$-bundle associated to $SM$ requiring that $(p_1,z_1^*)$ and $(L_A(p_1),z_1^*T(A))$ are equivalent, where $^*$ denotes the adjoint with respect to the inner product on $\bC^4$ and elements of $\bC^{4*}$ are understood as row vectors. Consequently, the dual pairing of $\bC^4$ and $\bC^{4*}$ extends in a well-defined way to a fibrewise dual pairing of $DM$ and $D^*M$.}

\definizione\label{testf}{Let $E$ be an arbitrary vector bundle over $M$. With $\cE(E)\doteq C^\infty(M,E)$, we denote the {\bf space of smooth sections} of $E$, endowed with the topology
induced by the family of seminorms
$$
\| f \|_{k,C} \doteq sup\{ | f^{(k)}(x) |, x \in C  \} \;, \qquad f\in \cE(E)\;,
$$
where $C$ is an arbitrary compact set and $\cdot\,^{(k)}$ denotes a derivative of $k$-th order. Furthermore, we introduce the {\bf space of smooth sections with compact support} $\cD(E)\doteq C_0^\infty(M,E)$, equipped with the topology induced by the family of 
seminorms
$$
\| f \|_{k} \doteq sup\{ | f^{(k)}(x) | \} \;, \qquad f\in \cD(E)\;.
$$
\\ 
\indent In the case $E=DM$, $E=D^*M$, we can define a {\bf global pairing} of $\cE(D^*M)$ and $\cD(DM)$ or $\cD(D^*M)$ and $\cE(DM)$ by integrating the local pairing induced by the inner product on $\bC^{4}$, {\it e.g.}, $$\langle f, g \rangle\doteq \int\limits_M d\mu(x) f(x)\left(g(x)\right)$$ for all $\quad f\in \cE(D^*M)$, $g\in \cD(DM)$.}

We are finally in the position to define the key object of our analysis:

\definizione\label{DirSpi}{A {\bf Dirac spinor} is a smooth global section of the Dirac bundle, {\it i.e.},
$\psi\in\cE(DM)$ or, equivalently, if we consider a local open neighbourhood $U$ of any point $x\in M$,
$\psi$ is (diffeomorphic to) a $4$-vector field $\psi_U$, {\it i.e.}, $\psi|_U\sim\psi_U:U\to\bC^4$, since the bundle trivialises as $DM|_U\sim 
U\times\bC^4$. Analogously, we call {\bf Dirac 
cospinor} a smooth global section of the dual Dirac bundle, namely, $\psi'\in\cE(D^*M)$ or, in 
a local neighbourhood $U$, $\psi'|_U\sim\psi'_U:U\to\bC^{4*}\sim\bC^4$.}

We would like to stress that our definition of Dirac spinor fields does not include any equation of motion.

It is quite safe to admit that the introduced geometric objects are rather technically complicated, and one
might wonder if the class of spacetimes admitting them is not rather restricted. Particularly, from a
physical point of view, one is interested in introducing spinors and fields in general as global objects.
There is thus a compelling need to understand which is really the set of backgrounds we can work with, and the
answer is surprising and reassuring at the same time. In fact, the following theorem can be proved (see
\cite{Borel} for the original proof and also \cite{Geroch} for a more physically oriented analysis and 
proof):

\teorema{\label{Withney}
A manifold $M$ admits a spin structure if and only if it has a vanishing second Stiefel-Whitney class or, in other
words, if the second de Rham\footnote{In the most general framework one should consider the second \v Cech
cohomology class, but this coincides with the de Rham one for differentiable manifolds. In this paper we
deal solely with spacetimes which, according to the definition stated at the beginning of the section, are
 differentiable.} cohomology class $H^2(M,\bZ_2)$ is trivial. Furthermore, if the manifold is four-dimensional, space and time-oriented, as well as globally hyperbolic, this requirement is automatically
satisfied.} 

This theorem is rather useful because globally hyperbolic spacetimes are the most interesting and natural 
class of manifolds whenever one deals with both  classical and quantum field theory on curved 
backgrounds. As a matter of fact, each of these spacetimes can be foliated as $\Sigma\times\bR$, being 
$\Sigma$ a smooth Cauchy surface \cite{Bernal}. Therefore, on these backgrounds one can state precisely the 
notion of initial value problem for the equations of motion, hence determining the classical dynamically 
allowed configurations of a field as the solution of a certain partial differential equation.

The property of globally hyperbolic, four-dimensional spacetimes $M$ which guarantees the existence of a spin structure is their parallelisability, {\it i.e.}, the fact that there always exists a global orthogonal frame on them. Consequently, $FM$ is a trivial bundle in that case, and this property extends to $SM$, $TM$, $T^*M$, $DM$, and $D^*M$ as well. We are thus in the position to introduce global frames of the latter four bundles.

\begin{enumerate}
\item Employing a global section $E$ of $SM$, we can define a {\bf spin frame} $\{E_A\}_{A=1\cdots4}$, {\it i.e.}, a set of four global sections of 
$DM$ as $E_A(x) \doteq [(E(x),z_A)]$, being $z_A$ the standard basis of $\bC^4$. Hence, any Dirac spinor $\psi$ can be decomposed as $\psi(x)=\psi^A(x)E_A(x)$ 
where now $\psi^A\in C^\infty(M)$. 
\item A {\bf dual spin frame} $\{E^B\}_{B=1,\cdots,4}$, {\it i.e.}, a set of four global sections of $D^*M$ can then be automatically
constructed out of the frame requiring $E_A(E^B)=\delta_A^B$. {\em From now on capital letter indices 
will refer to quantities expressed with respect to these bases.}
\item Exploiting definition \ref{spinstr}, we can project $E$ to $FM$, hence obtaining a global section $e\doteq \rho\circ E$ of $FM$. Employing this, we can define a {\bf Lorentz frame} $\{e_a\}_{a=0,\cdots, 3}$, {\it i.e.}, a set of four global sections of $TM$, by realising that $TM$ can be understood as the $\bR^4$-bundle associated to $FM$. Such a fibrewise basis is orthonormal in the sense that $g(e_a, e_b)=\eta_{ab}$, where $\eta_{ab}$ is the Minkowski metric, and it is often referred to as the
\emph{non-holonomic} basis of the base manifold; this is in sharp contrast with the standard (holonomic)
coordinate basis $\partial_\mu$ which is related to the basis $e_a$ by means of a basis change or, in other
words, by the matrix $e_a^\mu$ denoting the coefficients of $e_a$ in their expansion with respect to $\partial_\mu$.
\item Analogous to the spin case, one can now straightforwardly define a {\bf dual Lorentz frame} $\{e^b\}_{b=0,\cdots ,3}$ constructed out of $e_a$ as $e_a(e^b)=\delta_a^b$. {\em From now on, lower-case Roman letters will always refer to quantities expressed with respect to the
non-holonomic basis, whereas lower-case Greek ones will indicate those with respect to the holonomic basis. Non-holonomic indices will be ``raised'' and ``lowered'' with $\eta_{ab}$, while the same operations will be performed on the holonomic ones using $g_{\mu\nu}=g(\partial_\mu, \partial_\nu)$.}
\end{enumerate}

\noindent The most notable consequence of this new data is that we can decompose evert spinor-tensor
$$f\in\cE(\underbrace{TM\otimes\cdots\otimes TM}_i\otimes\underbrace{T^*M\otimes\cdots\otimes T^*M}_j\otimes
\underbrace{DM\otimes\cdots\otimes DM}_k\otimes\underbrace{D^*M\otimes\cdots\otimes D^*M}_l)$$
as follows
$$f=f_{b_1\cdots b_jB_1\cdots B_l}^{a_1\cdots a_iA_1\cdots A_k}e_{a_1}\otimes\cdots\otimes e_{a_i}\otimes e^{b_1}\otimes\cdots\otimes e^{b_j}
\otimes E_{A_1}\otimes\cdots\otimes E_{A_k}\otimes E^{B^1}\otimes\cdots\otimes E^{B^l}.$$
One could in principle certainly choose a different global section $E'$ of $SM$ and thus obtain different spin and Lorentz frames which are related to the previous ones by local spin and Lorentz transformations. On the level of coefficients, such a change of frames results in 
$$f_{b'_1,...,b'_j,B'_1,...,B'_l}^{\prime a'_1,...,a'_i,A'_1,...,A'_k}=\left(\Lambda^{-1}\right)^{a'_1}_{a_1}\cdots\left(\Lambda^{-1}\right)^{a'_i}_{a_i}\left(\widetilde\Lambda^{-1}\right)^{A'_1}_{A_1}\cdots\left(\widetilde\Lambda^{-1}\right)^{A'_k}_{A_k}
\Lambda^{b'_1}_{b_1}\cdots\lambda^{b'_j}_{b_j}\widetilde\Lambda^{B'_1}_{B_1}\cdots
\widetilde\Lambda^{B'_k}_{B_l}f_{b_1\cdots b_jB_1\cdots B_l}^{a_1\cdots a_iA_1\cdots A_k},$$
where $\widetilde\Lambda\in T\left(SL(2,\bC)\right)$, whereas $\Lambda=\Pi(\widetilde\Lambda)\in SO_0(3,1)$.

\subsection{On the dynamics of a classical Dirac field}
Since we have by now assured ourselves of the existence and well-posedness of the global structure of Dirac fields in a 
globally hyperbolic time and space-oriented manifold, we shall next proceed to introduce the natural evolution
operator out of which one can describe the classical dynamical content of our theory. It is imperative to
stress a sharp difference between the previous and this subsection; in the preceding discussion, all the 
introduced geometric structures have been somehow natural and intrinsic, {\it i.e.}, no special choice has been
performed with the due exception of the $D^{\left(\frac{1}{2},0\right)}\oplus D^{\left(0,\frac{1}{2}\right)}$
representation to define the Dirac bundle. Conversely, in the forthcoming discussion, some arbitrariness 
appears and we shall try to emphasise it to a potential reader, since we have to keep track of it to ensure that it does not play a 
distinguished role in the forthcoming discussion of the quantum fields. To wit, we are referring to the definition of the so-called $\gamma$-matrices. To obtain them, we can proceed as follows:

\definizione{Given $\bR^{p,q}$ endowed with the metric $\eta$ of signature $(p,q)$, we 
call {\bf Clifford algebra} $Cl(p,q)$ of $\bR^{p,q}$ the real associative algebra generated by the identity $I$ and an 
orthonormal basis of $\bR^{p,q}$ whose elements $\gamma_a$ with $a=1,...,p+q$ are subject to the 
relations
\beq
\left\{\gamma_a,\gamma_b\right\}\doteq\gamma_a\gamma_b+\gamma_b\gamma_a= 2\eta_{ab}I.
\eeq
Particularly, if $p=3$ and $q=1$ or vice-versa, one can refer to $Cl(3,1)$ or $Cl(1,3)$ as {\bf
Dirac-Clifford algebra}.}

\remark{It is a direct consequence of this definition that a basis for the Clifford algebra is given by the identity
and by all products $\gamma_{a_1}....\gamma_{a_n}$ with $a_1<...<a_n$ and $n\leq p+q$, which entails that 
$\dim(Cl(p,q))=2^{p+q}$. As a further important datum, we wish to underline that $Cl(p,q)$ is a
$\bZ_2$-graded algebra; this arises if we introduce the automorphism $\alpha:Cl(p,q)\to Cl(p,q)$ such that 
$\alpha(\gamma_a)=-\gamma_a$ for all possible $a$. Since $\alpha^2$ coincides with the identity map, we can 
always decompose:
$$Cl(p,q)=Cl^0(p,q)\oplus Cl^1(p,q),$$
where $Cl^i(p,q)=\left\{a\in Cl(p,q)\;|\;\alpha(a)=(-)^ia\right\}$. By direct inspection, one can realise
that $Cl^0(p,q)$ is the subalgebra of the full Clifford algebra generated by
products of even numbers of $\gamma_a$.}

The Dirac-Clifford algebra enjoys many relevant properties of great interest for our discussion. As a
first step, using {\em the periodicity theorem} 4.1 in \cite{Lawson}, one can prove per direct inspection  
that $Cl(1,3)_\bC\doteq Cl(1,3)\otimes\bC$ is isomorphic to the algebra $M(4,\bC)$ of $4\times 4$ complex 
matrices. This entails that it is natural to seek for a complex representation $T:Cl(1,3)_\bC\to
Hom(\bC^4,\bC^4)$ and, according to theorem 5.7, still in \cite{Lawson}, there is only one of these which is, up to
equivalence, irreducible. Its matrix form can be described by the matrices $\gamma_{aB} ^A$, which we choose
in such a way that $\left(\gamma^*_0\right)^{A}_{\;\;B}=-\gamma^A_{0B}$ whereas $\left(\gamma^*_a\right)^{A}_{\;\;B}=
\gamma^A_{aB}$ for  $a=1,2,3$, {\it i.e.},
\beq\label{gammas}
\gamma_0=i\left(\begin{array}{cc}
I_2 & 0\\
0 & -I_2
\end{array}\right),\quad\gamma_a=i\left(\begin{array}{cc}
0 & \sigma_a\\
-\sigma_a & 0
\end{array}\right),
\eeq
where $a=1,...,3$ and $\sigma_a$ is a Pauli matrix, whereas $I_n$ denotes the $n\times n$-identity matrix.
Furthermore, these matrices, independently of the choice in \eqref{gammas}, are the so-called
{\em $\gamma$-matrices} and they can always be interpreted as the coefficients of a global tensor $\gamma\in\cE(T
^*M\otimes DM\otimes D^*M)$ admitting the following expansion:
\beq\label{glueing}
\gamma=\gamma_{aB}^A e^a\otimes E_A\otimes E^B.
\eeq
This identity yields that, if we use the tetrad coefficients $e^a_\mu$ to define $\gamma_\mu\doteq\gamma_a e^a_\mu$, we 
recover the standard anticommutation relations for the $\gamma$-matrices in curved backgrounds
\beq\label{gammamu}
\left\{\gamma_\mu,\gamma_\nu\right\}=2g_{\mu\nu}I_4\;.
\eeq
This is the net effect of changing from a non-holonomic to a holonomic basis.

Let us stress at this point that we have chosen a specific representation of the Dirac-Clifford algebra out of the possible equivalent ones and indeed this is the arising arbitrariness announced at the beginning of this subsection. The choice of different representations of the Dirac-Clifford algebra can be shown to lead to quantum field theories which are equivalent up to gauge transformations \cite{Sanders} and, moreover, if one restricts to the quantum observables, the choice of a representation becomes even irrelevant, as we will discuss in the next section. 

A second interesting property of the Clifford algebra arises from the realisation that it contains the spin group (see \S 2 of \cite{Lawson}) and, most notably, $Spin(3,1)\subset Cl_0(3,1)$. It is thus
natural to wonder how the above introduced representation yielding the $\gamma$-matrices restricts to the
spin group, and the answer to this query comes from proposition 5.15, still in \cite{Lawson}, which guarantees 
us that each irreducible complex representation of the Clifford algebra on a vector space can be restricted 
to the sum of two inequivalent irreducible representations of the spin group. In the case under 
analysis, one can by direct inspection realise that the restriction of $T$ to $Spin(3,1)$ yields the sum of 
two non-equivalent irreducible representations, which, on $SL(2,\bC)\sim Spin_0(3,1)$, coincide with the previously mentioned
$D^{\left(\frac{1}{2},0\right)}\oplus D^{\left(0,\frac{1}{2}\right)}$ representation.

\noindent According to our analysis, for any vector field $v\in\cE(TM)$, we can meaningfully introduce
$$\displaystyle{\not}{v}\doteq v^a\gamma_a,$$
which is an element of $\cE(DM\otimes D^*M
)$, such that its coefficients $v^A_{\;\;B}$ form a $4\times 4$ complex matrix. Notice that, from now on, we will \underline{not specify} explicitly when we shall deal with
abstract Dirac-Clifford algebra elements $\ga_a$ or with their matrix representations. It is understood that, whenever we either  contract
$\gamma$-matrices with a vector field or these matrices are applied to a vector in $\bC^4$, we refer to the
latter case.

The last ingredient we need to specify the dynamics of Dirac fields is a parallel transport on the Dirac bundle. The grand strategy is rather
simple, namely, we introduce the standard metric connection, interpret it on the frame bundle and
eventually lift it to both the spin and the Dirac bundle:

\definizione\label{covd}{Let $\om:FM\to T^*FM\otimes {\mathfrak o(3,1)}$ denote the connection 1-form of the unique {\bf Levi-Civita
connection} on $FM$. It induces the standard Levi-Civita connection on $TM$ (and vice versa) which can be expressed as the covariant derivative $$\nabla:\cE(TM)\to\cE(TM\otimes T^*M),\quad\nabla e_a=\Gamma^b_{ac}e_b\otimes e^c,$$ where the {\em connection coefficients} $\Gamma^b_{ac}$ of the Levi-Civita connection are specified by $$\Gamma^b_{ac}\doteq e^b\left((\om\circ e)\left[e_*\circ e_a\right]e_c\right),$$ 
with $e_*: T^*M\to T^*FM$ denoting the push-forward of $e$ in the sense of cotangent vectors.\\
\indent The pull-back $\Om\doteq (d\Pi)^{-1}\circ \rho^*\circ\om$ of $\om$ to $SM$, with $d\Pi: {\mathfrak sl}(2,\bC)\to{\mathfrak o(3,1)}$ denoting the derivative of the covering $\Pi$ at the identity, defines the {\bf spin connection}, which by the definition of $DM$ as a bundle associated to $SM$ can be specified as a covariant derivative
$$\nabla:\cE(DM)\to\cE(DM\otimes T^*M),\quad \nabla E_A=\sigma^B_{aA}e^a\otimes E_B,$$ where the spin connection coefficients 
are given by \beq\label{sigmaexp}\sigma^B_{aA}\doteq E^B\left((\Om\circ E)\left[E_*\circ e_a\right]E_A\right)\eeq and $E_*: T^*M\to T^*SM$ denotes the push-forward of $E$ in the sense of cotangent vectors.}

\noindent We will soon prove that the coefficients $\sigma^B_{aA}$ can be expressed in a simple way by means of both the coefficients $\Gamma^b_{ac}$ and the $\gamma$-matrices. Before we tackle such task, let us stress
that the covariant derivatives $\nabla$ from definition \ref{covd} can be straightforwardly extended to any spinor-tensor
$$f\in\cE(\underbrace{TM\otimes\cdots\otimes TM}_i\otimes\underbrace{T^*M\otimes\cdots\otimes T^*M}_j\otimes
\underbrace{DM\otimes\cdots\otimes DM}_k\otimes\underbrace{D^*M\otimes\cdots\otimes D^*M}_l)$$ by defining

\begin{gather}\nabla:\cE(\underbrace{TM\otimes\cdots\otimes TM}_i\otimes\underbrace{T^*M\otimes\cdots\otimes T^*M}_j\otimes
\underbrace{DM\otimes\cdots\otimes DM}_k\otimes\underbrace{D^*M\otimes\cdots\otimes D^*M}_l)\to\notag\\
\to\cE(T^*M\otimes\underbrace{TM\otimes\cdots\otimes TM}_i\otimes\underbrace{T^*M\otimes\cdots\otimes T^*M}_j\otimes
\underbrace{DM\otimes\cdots\otimes DM}_k\otimes\underbrace{D^*M\otimes\cdots\otimes D^*M}_l).\label{covderi}
\end{gather}
At a level of components and, for notational simplicity, only in the case $i=j=l=k=1$, \eqref{covderi} reads
\begin{gather*}
\nabla f=e^c\nabla_c\left( f_{bB}^{aA}e_{a}\otimes e^{b}\otimes E_{A}\otimes
E^{B}\right)=\\
\left[\partial_c f_{bB}^{aA}-\Gamma_{cb}^d f_{dB}^{aA}+\Gamma_{cd}^a f_{bB}^{dA}-\sigma_{cC}^{A}f_{bB}^{aC}+\sigma_{cB}^C f_{bC}^{aA}\right]e^c\otimes e_{a}\otimes e^{b}\otimes E_{A}
\otimes E^{B}.
\end{gather*}

As promised, we shall now give an explicit expression for the $\sigma$-coefficients. Although a 
demonstration of the next lemma is already present in \cite{Lichnerowicz} (see also chapter 13 of \cite{Wald}), we shall prove it again due to ubiquitous sign subtleties arising from the choice of the metric signature and possibly leading to 
confusions when comparing with the literature, \cite{Dimock} in particular.

\lemma\label{covdev}{The connection coefficients of the spin connection can be expressed as
\beq\label{covder}
\sigma^B_{aA}=\frac{1}{4}\Gamma^b_{ad}\gamma_{bC}^B\gamma^{dC}_{\;\;\;A}\,.
\eeq}

\proof{The strategy of the proof is the following: we first derive an explicit expression for the double covering homomorphism $\Pi: Spin_0(3,1)\sim SL(2,\bC)\twoheadrightarrow SO_0(3,1)$. From this we obtain an expression for its derivative at the identity $d\Pi: {\mathfrak sl}(2,\bC)\to{\mathfrak o(3,1)}$, which, inserted in (\ref{sigmaexp}), yields the wished-for result.\\\indent
Let us thus recall that $Spin_0(3,1)$ can be understood as a subgroup of $Cl(3,1)$. This 
entails that, for any $\widetilde\Lambda\in SL(2,\bC)$, we can define the adjoint action
$$Ad_{\widetilde\Lambda}(\gamma_a)\doteq{\widetilde\Lambda}\gamma_a{\widetilde\Lambda}^{-1}.$$ Being $SL(2,\bC)$ a finite cover of $SO_0(3,1)$, applying an induction-reduction mechanism, we know it exists a
representation $T$ of $SO_0(3,1)$ such that, setting $\La\doteq\Pi(\widetilde\Lambda)$, we have 
\beq\label{aux}
T(\Lambda)\gamma_a=Ad_{\widetilde\Lambda}(\gamma_a).
\eeq
Since we are dealing with Lie groups, we recall that all finite-dimensional representations are matrix
representations, {\it i.e.}, we can simply write $T(\La)\gamma_a\doteq \La^b_{\;\;a}\gamma_b$. Hence,  ${\widetilde\Lambda}\gamma_a{\widetilde\Lambda}^{-1}=\Lambda^b_{\;\;a}\gamma_b$, where we notice the invariance of the left hand side under the
$\mathbb{Z}_2$-action sending $\widetilde\Lambda\to\pm\widetilde\Lambda$, as one could have expected, being $SL(2,\bC)$ the double
cover of $SO_0(3,1)$.\\\indent
Let us now take an arbitrary differentiable path $t\mapsto\widetilde\Lambda(t)$ in $SL(2,\bC)$ whose projection on $SO_0(3,1)$ is
differentiable; the following identity holds 
$$\widetilde\Lambda(t)\gamma_a\widetilde\Lambda(t)^{-1}=\Lambda(t)^b_{\;\;a}\gamma_b.$$
If we derive with respect to $t$, an identity between algebra representations arises, namely,
\begin{gather}
\frac{d\widetilde\Lambda(t)}{dt}\gamma_a\widetilde\Lambda(t)^{-1}+\widetilde\Lambda(t)\gamma_a\frac{d\widetilde\Lambda(t)^{-1}}{dt}=\left(\frac{d\Lambda(t)
}{dt}\right)^b_{\;\;a}\gamma_b\notag\\\label{quarterone}
\Leftrightarrow\quad \frac{d\widetilde\Lambda(t)}{dt}\gamma_a\widetilde\Lambda(t)^{-1}-\widetilde\Lambda(t)\gamma_a\widetilde\Lambda(t)^{-1}\frac{d\widetilde\Lambda(t)}{dt}
\widetilde\Lambda(t)^{-1}=
\left(\frac{d\Lambda(t)}{dt}\right)^b_{\;\;a}\gamma_b,
\end{gather}
where we have exploited that the derivation of $\widetilde\Lambda(t)\widetilde\Lambda(t)^{-1}=I$ yields
$$\frac{d\widetilde\Lambda(t)}{dt}\widetilde\Lambda(t)^{-1}=-\widetilde\Lambda(t)\frac{d\widetilde\Lambda(t)^{-1}}{dt}.$$
Applying the adjoint action of $\widetilde\Lambda(t)^{-1}$ to (\ref{quarterone}), we end up with
$$\widetilde\Lambda(t)^{-1}\frac{d\widetilde\Lambda(t)}{dt}\gamma_a-\gamma_a\widetilde\Lambda(t)^{-1}\frac{d\widetilde\Lambda(t)}{dt}=Ad_{\widetilde\Lambda(t)^{-1}}\left[\left(\frac{d\Lambda(t)}{dt}
\right)^b_{\;\;a}\gamma_b\right].$$
We use \eqref{aux} as well as the basic property of a representation, namely, $Ad(\widetilde\Lambda^{-1})\equiv Ad(
\widetilde\Lambda)^{-1}$, to derive
$$\widetilde\Lambda(t)^{-1}\frac{d\widetilde\Lambda(t)}{dt}\gamma_a-\gamma_a\widetilde\Lambda(t)^{-1}\frac{d\widetilde\Lambda(t)}{dt}=
\left(\Lambda^{-1}\right)^{c}_{\;\;a}\left(\frac{d\Lambda(t)}{dt}\right)^b_{\;\;c}\gamma_b=\left(\Lambda^{-1}\frac{d\Lambda(t)}{dt}\right)^b_{\;\;a}\gamma_b.$$
If we call $\lambda\doteq\widetilde\Lambda(t)^{-1}\frac{d\widetilde\Lambda(t)}{dt}$ and $\mu\doteq \Lambda(t)^{-1}\frac{d\Lambda(t)}{dt}$, it 
holds $\lambda\gamma_a-\gamma_a\lambda=\mu^b_{\;\;a}\gamma_b$. Let us multiply this identity on the right with  
$\gamma_a$ and, if we bear in mind that $\gamma^a\gamma_a=\eta^{ab}\gamma_a\gamma_b=\eta^{ab}\eta_{ab}=4$, 
we end up with
\beq\label{inm}
4\lambda-\gamma_a\lambda\gamma^a=\mu_{ab}\gamma^a\gamma^b.
\eeq
Taking into account the antisymmetry of $\mu\in{\mathfrak o(3,1)}$ and the identity $\gamma^a\gamma^{[b}\gamma^{c]}\gamma_a=0$, a possible solution of \eqref{inm} is
$$\lambda=\frac{1}{4}\mu_{ab}\gamma^a\gamma^b.$$
The uniqueness of this solution is not guaranteed, since, being the left hand side of \eqref{inm} linear in
$\lambda$, we have merely found a particular solution, and we are free to add any further
solution of the homogeneous counterpart, {\it i.e.}, any $\overline\lambda$ such that $4\overline\lambda-
\gamma_a\overline\lambda\gamma^a=0$. This implies that the commutator $[\gamma_a,\overline\lambda]=0$, and
Schur's lemma (see th. 4.26 in \cite{Hall}) entails that $\overline\lambda=kI$, being $I$ the identity.\\\indent
The value of $k$ can be unambiguously determined if we notice that, according to our previous construction,
$\overline\lambda$ is an element in the algebra of $SL(2,\bC)$, which consists of matrices with vanishing trace.
Therefore, if we impose $Tr(\overline\lambda)=Tr(kI)=0$, the only possibility is $k=0$. \\\indent
Since the differentiable path chosen in the proof has been arbitrary, we have now proven the explicit form of $d\Pi$ in terms of its inverse, namely,  $$(d\Pi)^{-1}: {\mathfrak o(3,1)}\to{\mathfrak sl}(2,\bC),\quad  (d\Pi)^{-1}(\mu_{ab})=\frac{1}{4}\mu_{ab}\gamma^a\gamma^b\;\;\forall \mu_{ab}\in{\mathfrak o(3,1)}.$$
\indent Remembering the definition of $\Om$, recalling $e=\rho\circ E$, and inserting the expression of $(d\Pi)^{-1}$ into (\ref{sigmaexp}), we finally obtain $$\sigma^B_{aA}=\frac{1}{4}\Gamma^b_{ad}\gamma_{bC}^B\gamma^{dC}_{\;\;\;A}.$$}

\lemma\label{DC}{The Dirac-Clifford $\gamma$-matrices are covariantly constant, {\it i.e.}, $\nabla\gamma=0$.
}

\proof{This is a straightforward calculation, once a subtlety has been clarified: the matrices
$\gamma$ are constructed as the matrix form of an irreducible representation of the Clifford algebra and then subsequently
glued to each point of the underlying base manifold $M$ via \eqref{glueing}. This prescription entails 
that $\partial_a\gamma_b=0$ for any $a,b=0,...,3$.\\ 
\indent We can now compute
$$\nabla\gamma=e^b\nabla_{b}\left(\gamma_{aB}^A e^a\otimes E_A\otimes E^B\right),$$
which, exploiting definition \ref{covd}, becomes
\begin{gather*}
\nabla\gamma=e^b\left[\partial_b\gamma_{aB}^A e^a\otimes E_A\otimes E^B+\gamma_{aB}^A\left(\Gamma^{a}_{bc}e^c
\otimes E_A\otimes E^B+\right.\right.\\
\left.\left.-\sigma_{bA}^C e^a\otimes E_C\otimes E^B+\sigma_{bC}^B e^a\otimes E_A\otimes E^C\right)\right].
\end{gather*}
If we apply formula \eqref{covderi} to sections of $T^*M\otimes DM\otimes D^*M$ and consider that the matrix elements are constant functions, the lemma is proved
out of direct substitution.}

\noindent We are ready, at last, to describe the dynamically allowed configurations of a spinor field in a
curved background. Particularly, if we now understand multiplication with the $\gamma$-matrices to act from the left on spinors and from the right on cospinors, we shall call
dynamically allowed a Dirac (co)spinor $\psi^{(\prime)}$ which satisfies the {\bf Dirac equation}\footnote{
In Minkowski spacetime, the Dirac equation is most notable for having a purely imaginary
coefficient $i$ in front of the Dirac operator. Here, such a number does not appear and the underlying
reason is rooted in the employed convention for both the metric signature and the sign of the defining
anticommutation relations for the Clifford algebra. To wit, $i$ is absorbed in the definition of the
$\gamma$-matrices, as stated in (\ref{gammas}).}
\begin{gather}\label{Dirac}
D\psi\doteq\left(-\displaystyle{\not}\nabla+m\right)\psi=0,\\
D'\psi'\doteq\left(\displaystyle{\not}\nabla+m\right)\psi'=0, \label{Dirac2}
\end{gather}
where $\psi\in\cE(DM)$, whereas $\psi'\in\cE(D^*M)$.

\subsection{The Cauchy problem and the fundamental solutions}

In this subsection we shall discuss the classical initial value problem for Dirac fields. As it is well-known from flat spacetimes, a solution of the Dirac equation is usually related to a solution of a hyperbolic differential equation.

\lemma\label{waveeq}{Every solution $\psi$ of \eqref{Dirac} is also a solution of the spinorial Klein-Gordon equation 
\beq\label{waveD}
P\psi\doteq\left(\nabla_a\nabla^a-\frac{R}{4}-m^2\right)\psi=0,
\eeq
where $R$ is the scalar curvature of $(M,g)$. A similar
statement holds for each cospinor $\psi'$, solution of \eqref{Dirac2}.}

\proof{Let us take a solution of \eqref{Dirac} and let us multiply it with
$D'$. We end up with
$$D'D\psi=(\dnot\nabla+m)(-\dnot\nabla+m)\psi=(-\dnot\nabla\dnot\nabla+m^2)\psi=0,$$
which means that we need to prove that $\dnot\nabla^2=\nabla_a\nabla^a-\frac{R}{4}$. To this avail, let us write
$$\dnot\nabla^2=\gamma^a\nabla_a\gamma^b\nabla_b=\gamma^a\gamma^b\nabla_a\nabla_b=\gamma^{(a}
\gamma^{b)}\nabla_a\nabla_b+\gamma^{[a}\gamma^{b]}\nabla_a\nabla_b,$$
where in the second equality we have exploited lemma \ref{DC}, whereas in the third one we split the
expression in its symmetric and antisymmetric part. Since $\gamma^{(a}\gamma^{b)}=\frac12 \{\gamma^{a},\gamma^{b}\}$ is equal to the metric times the identity, it holds
\beq\label{kleingordonfirst}
\dnot\nabla^2=\nabla_a\nabla^a+\gamma^{[a}\gamma^{b]}\nabla_a\nabla_b=\nabla_a\nabla^a+\gamma^a\gamma^b\nabla_{[a}\nabla_{b]}=\nabla_a\nabla^a+\frac{1}{2}\gamma^a\gamma^b C_{ab},
\eeq
with $C_{ab}$ denoting the curvature tensor of the spin connection.\\\indent
Let us briefly state some properties of $C_{ab}$, which are examined in more detail in appendix \ref{NC}: firstly, from lemma \ref{covdev} one can infer that $$C_{ab}=\frac14 R_{abcd}\ga^c\ga^d.$$ Employing the Clifford relations and the symmetry properties of the Riemann tensor $R_{abcd}$, it is straightforward to show that $$\ga^a\ga^b C_{ab}=-\ga^b\ga^a C_{ab}=-\frac 12 \ga^b\ga^a R_{ab}=-\frac R 2.$$
Inserting this into (\ref{kleingordonfirst}), we finally obtain
$$\dnot\nabla^2=\nabla_a\nabla^a-\frac{R}{4}$$ and thus $$D'D=DD'=-P=-\nabla_a\nabla^a+\frac{R}{4}+m^2.$$}

\remarks{One should notice that the above lemma can be seen as a particular application of
Weizenb\"{o}ck's formula \cite{Lawson}.\\
\indent It is furthermore worth noting that $\nabla_a\nabla^a$ in the above expression is \underline{not} diagonal in the spinor indices, and thus not $\square$ times the identity. Its principal symbol $g^{\mu\nu}k_\mu k_\nu$, however, is indeed diagonal and even of metric type.\\
\indent Let us also stress that, in sharp contrast to the scalar Klein-Gordon equation in four spacetime dimensions, there is no freedom to select a coupling for Dirac fields to the scalar curvature since this is universally fixed to $\frac14$.
Furthermore, for Dirac fields, such factor corresponds to the conformal coupling whereas, for the scalar case, the latter is  $\frac16$.}

The introduction of the Dirac operator and the discussion of its main properties allow us to state and to
prove the main theorem related to the classical dynamical behaviour of spinors on curved backgrounds:

\teorema\label{secondord}{Let $(M,g)$ be a four-dimensional globally hyperbolic oriented and time oriented spacetime, such that $M$ is diffeomorphic to $\Sigma\times\bR$, with $\Si$ a three-dimensional Riemannian manifold. Let us furthermore call $\iota: \Si\hookrightarrow M$ a smooth embedding of $\Si$ into $M$, such that $\iota(\Si)$ is a Cauchy surface of $M$. If we refer to $D\Sigma$ as the bundle constructed out of the pull-back of $DM$ via $\iota$, then the following Cauchy problem admits a unique solution
\beq\label{CauchyDir}
\left\{\begin{array}{l}
D\psi=0\\
\iota^*\psi\equiv\psi_0
\end{array}\right.,
\eeq
where $\psi\in\cE(DM)$ and $\psi_0\in\cD(D\Si)$.}

\proof{ 
For notational simplicity we will omit the embedding $\iota$ and just write $\psi|_\Si$ instead of $\iota^* \psi$ for any $\psi\in\cE(DM)$.\\\indent
According to lemma \ref{waveeq}, each solution of the
Dirac equation also solves a spinorial counterpart of the Klein-Gordon equation, namely, $P\psi=\left(
\square-\frac{R}{4}-m^2\right)\psi=0$ where $P$ has metric principal symbol and is thus a normally hyperbolic differential operator.\\\indent Hence, we can invoke the results on hyperbolic partial differential equations (see \cite{Bar} and theorem
3.2.11 in particular), which guarantee us that any Cauchy problem for compactly supported initial data for the 
the considered partial differential equation admits a unique smooth solution supported in the causal past and 
future of the initial datum. The only problem left is to give a prescription on how to switch from a Cauchy 
problem for the Dirac equation, hence with only one given initial datum, to one for a second order 
hyperbolic partial differential equation, where two data on the Cauchy surface have to be prescribed. To solve 
this dilemma, let us not deal with \eqref{CauchyDir}, but with the following system:
$$ \left\{\begin{array}{l}
\left(\square-\frac{R}{4}-m^2\right)u=0\\
u|_\Si=0,\quad -\frac{\partial u}{\partial n}|_\Si\equiv
\displaystyle{\not}{n}\psi_0
\end{array}\right., $$
where $u\in\cE(DM)$ and $n$ denotes the vector field normal to $\Sigma$ such that $\eta^{ab}n_a n_b=-1$. As stated before, such a system admits a unique global smooth solution $\tilde u$ and, hence, let us introduce the smooth section $\widetilde\psi\doteq D'\tilde u$. It is immediate to see that $\widetilde\psi$ is a solution of the Dirac equation $D\widetilde\psi=0$. Since $\widetilde\psi$ is smooth, it can be directly evaluated on the Cauchy surface $\Si$ where $\widetilde\psi|_\Sigma=D'\tilde u|_\Sigma=\left(\displaystyle{\not}\nabla+m\right)\tilde u|_\Sigma$. Inserting the initial condition for $\tilde u$, the
second term vanishes, whereas the first reads
$$\gamma^a\nabla_a \tilde u|_\Sigma=\gamma^a n_a\left.\frac{\partial \tilde u}{\partial n}\right|_\Sigma=
-\gamma^a n_a\gamma^b n_b\psi_0=-\eta^{ab}n_a n_b\psi_0=\psi_0.$$
Hence, the section $\widetilde\psi$ constitutes a unique solution of the Cauchy problem in the
thesis of the theorem.}

Since our aim is to deal with quantum field theory over curved backgrounds in the long run, it is more useful to prove 
not only the existence and uniqueness of the solution of a Cauchy problem, but also the existence of 
the so-called fundamental solution, which is the last relevant theorem of the classical theory that we shall need:
\teorema\label{fund}{The Dirac operator admits unique advanced ($^-$) and retarded ($^+$) fundamental solutions, {\it i.e.}, continuous linear maps $S^\pm
:\cD(DM)\to \cE(DM)$ fulfilling $DS^\pm=I=S^\pm D$. These maps are determined by their support properties $$supp(S^\pm f)\subset
J^\pm(supp\; f), \quad\forall f\in \cD(DM),$$ with $J^\pm(U)$ denoting the causal future/past of the set $U$. Similarly, there exist unique advanced and retarded fundamental solutions $S_*^\pm: \cD(D^*M)\to \cE(D^*M)$ of $D'$. 
}

\proof{The strategy will be similar to the one employed in the proof of the existence of the
solution of the Cauchy problem. Thus, let us start from $P=-D'D=-DD'$; since this is a normally hyperbolic differential operator, we already know (see \cite{Bar, Hormander3})
that it exists a unique advanced - say $E^+$ - and a unique retarded - say $E^-$ - fundamental solution for $P$ on $\cD(DM)$. Hence, for any $f\in\cD(DM)$, we know that $$PE^\pm=I=E^\pm P\quad\textrm{and}\quad supp(E^\pm f)\subset
J^\pm(supp\; f).$$
We can now define $S^\pm\doteq-D'E^\pm$ which, per direct inspection, satisfies
$DS^\pm=I$ and is furthermore continuous and has the correct support properties, since the application of $D'$ preserves these features. In the same way, we can construct advanced and retarded right fundamental solutions for the dual Dirac operator as
$S_*^\pm\doteq-DE_*^\pm$, with $E_*^\pm$ being the fundamental solutions of $P$ on $\cD(D^*M)$.\\\indent
The next step consists of proving that the right fundamental solution is ``also a left one'' and we only show this for $S^\pm$, since the proof for $S_*^\pm$ is analogous. Consider any $h\in\cD(D^*M)$ and any $f\in\cD(DM)$. Since $D f\in \cD(DM)$, we end up with
\begin{gather*} 
\langle h, S^\pm Df\rangle=\langle D' S_*^\mp h,S^\pm Df\rangle=\langle S_*^\mp h,DS^\pm D f\rangle=\\
=\langle S_*^\mp h,D f\rangle=\langle D' S_*^\mp h,f\rangle=\langle h,f \rangle,
\end{gather*}
where all expressions are well-defined since supp$S^\pm f\,\cap$ supp$S_*^\mp h$ is compact due to global hyperbolicity of $M$.\\\indent
It remains to be shown that the fundamental solutions are unique and again we only prove this for $S^\pm$ here. To this end, let us take any $h\in\cD(M,D^*M)$ and any
$f\in\cD(M,DM)$. Suppose two different sets of fundamental solutions, say $S^\pm$ and $\widetilde S^\pm$, exist. Starting from 
$$0=\langle Ih-Ih,f\rangle=\langle(S^\pm-\widetilde S^{\pm})D h, f\rangle,$$ uniqueness of the right fundamental solutions follows from the non-degeneracy of $\langle\,,\,\rangle$, while uniqueness in the sense of left fundamental solutions can be deduced in a similar way.}

In analogy with the scalar case, we shall from now on call $S=S^+-S^-$ the {\bf causal propagator} for the 
Dirac operator $D$ and $S_*= S_*^+-S_*^-$ the causal propagator for $D'$.

To conclude the section, we wish to underline that, up to this point, we have basically considered the Dirac spinor and cospinor fields as completely distinct objects. As it is usually done in Minkowski space, however, we can define a well-behaved Dirac conjugation map mapping spinors into cospinors and vice versa. Furthermore, this conjugation turns out to map any dynamically allowed configurations into another one.

\definizione{\label{Diracconj}We call {\bf Dirac conjugation matrix} the unique matrix $\beta\in SL(4,\bC)$ such that 
$$\beta^*=\beta,\quad\gamma_a^*=-\beta\gamma_a\beta^{-1}\qquad\forall a=0,...,3,$$
and furthermore $i\beta n^a\gamma_a>0$, being $n$ timelike and future-directed. \\\indent Starting from this object, we
can define {\bf Dirac conjugation maps}:
$$\cdot^\dagger:\cE(DM)\longrightarrow\cE(D^*M),\quad f^\dagger \doteq f^*\be,$$ 
$$\cdot^\dagger:\cE(D^*M)\longrightarrow\cE(DM),\quad h^\dagger \doteq \be^{-1}h^*,$$
where $^*$ denotes the adjoint with respect to the Hermitian inner product on $\bC^4$.
}

\remarks{$\be$ is only unique, once a representation of the Dirac-Clifford algebra has been chosen. A direct inspection of the above identities shows that, with the definition of $\gamma$-matrices as
in \eqref{gammas}, $\beta=-i\gamma_0$ and thus $\be=\be^{-1}$.\\
\indent It furthermore follows in general that applying the Dirac conjugation twice gives the identity and that, as already anticipated,  $\cdot^\dagger$ preserves the Dirac equations in the sense that $D'f^\dagger=(Df)^\dagger$, $Dh^\dagger=(D'h)^\dagger$ for any $f\in\cE(DM)$, $h\in\cE(D^*M)$.
}

\section{Dirac fields: quantum point of view}

The aim of this section is twofold: on the one hand, we shall discuss the already available formulation of a quantum theory for Dirac fields on a curved background while, on the other hand, we shall show for the first time that the notion of Wick polynomials can be coherently introduced also in this scenario, giving rise to an enlarged algebra of observables. Particularly, we shall point out how all these topics fit into the framework of the locally covariant formulation of quantum field theory.

To achieve our goal, we shall refer to an earlier work due to Araki \cite{Araki70}, though we shall also profit from 
\cite{Dimock, Koehler, Hollands, Kratzert, SahlmannVerch, VerchSpin, Fewster, Sanders}. 

\subsection{The local algebras of fields and observables}\label{laqft}
Although the first paper formulating a quantum theory of Dirac fields on curved spacetimes in the algebraic framework is \cite{Dimock}, its underlying approach is slightly different from the one we shall employ, albeit it is ultimately fully equivalent. Particularly, in the aforementioned paper, the quantisation scheme calls for the choice of both a Cauchy surface $\Sigma$ and initial data on $\Sigma$ as building blocks of the quantum theory. We shall not dwell into the details of this method since we reckon that, in the spirit of local covariance, it is not best suited for our later purposes.

Although the standard paradigm in particle physics calls for the treatment of particles and antiparticles as distinct, albeit related objects, in this paper we shall, as it has been done by most of the authors mentioned in the introduction of this section, bear in mind the lessons from \cite{Araki70} and, thus, we shall consider spinors and cospinors as part of a single entity, since it will turn out to be more convenient for our later purposes.

On a practical ground, the building blocks of our discussion will be three. The first one arises out of the direct sum $DM\oplus D^*M$, namely, $\cD\doteq\cD(M)\doteq\mD(DM\oplus D^*M)$, the space of compactly supported smooth sections of $DM\oplus D^*M$ with the usual topology, {\it i.e.}, the one induced by the family of seminorms
$$
\|f\|_k \doteq \sup{| f^{(k)}(x) |}\;,\qquad  f\in \cD\;, 
$$
while the second is the map
$\Ga:\mD\to\mD$ such that
\beq\label{bidagger}
\Ga(f \oplus h)= h^\dagger\oplus f^\dagger\quad\forall f\oplus h\in\mD,
\eeq
being $\cdot^\dagger$ the Dirac conjugation introduced in definition \ref{Diracconj}.
Furthermore, in order to eventually impose the anticommutation relations, we need a third datum, namely, a sesquilinear form on $\mD^2\doteq\mD\left((DM\oplus D^*M)^{\boxtimes 2}\right)$: let $f=f_1\oplus f_2$ and $h=h_1\oplus h_2$ be two elements of $\mD$. Then $(\cdot,\cdot):\mD^2\to \bC$ is defined as:
\beq\label{sesqui}
(f,h)\doteq -i \langle f_1^\dagger,S h_1 \rangle +i \langle S_* h_2, f_2^\dagger \rangle,  
\eeq
which is positive semidefinite, as one can infer with minor modifications either from lemma 4.2.4 in \cite{Sanders} or, with little more effort, from proposition 1.1 in \cite{Dimock}; let us also note that $(\Ga f, \Ga h)=(h,f)$. Furthermore, one can straightforwardly show that, using all the afore introduced tools, we can define the following algebra:

\definizione\label{fieldal}{We call {\bf algebra of fields} the unital $*-$algebra $\cF(M,g)$ generated by the identity and the abstract elements $B(f)$ with $f\in\mD$ satisfying the following requirements:
\begin{enumerate}
\item[i)] the map $f\mapsto B(f)$ is linear,
\item[ii)] $B(D f_1\oplus D' f_2)=0$ for all $f_1\oplus f_2\in\mD$,
\item[iii)] $B(\Ga f)= B(f)^*$, for all $f\in\mD$ and with $\Ga$ defined as in \eqref{bidagger},
\item[iv)] $\{B(f)^*,B(h)\}\doteq B(f)^*B(h)+B(h)B(f)^*=(f,h)$\;, where the right hand side is given by \eqref{sesqui}.
\end{enumerate}}

In order to convince a potential reader that we are employing a sensible definition, let us first discuss some general properties.

\remark{
It is possible to recover the standard notion of spinor and cospinor quantum field starting from the $B$-generators as follows:
\begin{itemize}
\item the spinor arises as $\quad\psi(h)\doteq B(0\oplus h)$.
\item the cospinor is given by $\quad\psi^\dagger(f)\doteq B(f\oplus 0)$, 
\end{itemize}
Particularly, to be convinced of the self-consistency of such statement, one should notice that the spinor and cospinor fields are related due to property iii) and they respectively satisfy the Dirac and the dual Dirac equation of motion in the distributional sense thanks to ii). Finally, it is iv) which corresponds to the usual anticommutation relations between $\psi$ and $\psi^\dagger$, namely, $$\{\psi(h),\psi^\dagger(f)\}=-i\langle h, Sf\rangle,\quad \{\psi(h_1),\psi(h_2)\}=\{\psi^\dagger(f_1),\psi^\dagger(f_2)\}=0.$$\vskip -1pt}
In order to better grasp the structure of $\cF(M,g)$, one should realise that it is nothing but a topological $*$-algebra; this can be fully understood starting from an {\it a priori} different, albeit ultimately equivalent, perspective, namely, the so called {\it Borchers-Uhlmann algebra} for Dirac fields, which is explicitly discussed in \cite{Sanders}. It is constructed as the following quotient
\beq\label{BorUhl}
\mF(M,g)\doteq\left(\bigoplus_{n=0}^\infty\mD^n\right)/\cI 
\eeq
where $\mD^n\doteq\mD\left((DM\oplus DM^*)^{\boxtimes n}\right)$ denotes the compactly supported sections of the $n$-fold outer tensor product of $DM\oplus D^*M$ and $\cD^0\doteq \bC$, whereas $\cI$ is the closed $*$-ideal which arises out of the relations (ii), (iii), and (iv) in definition \ref{fieldal}. It is hence generated by elements of the form $D f_1\oplus D' f_2$ with $f_1\oplus f_2\in\mD$, by those of the form $(f_1\oplus f_2)^* - \Ga(f_1\oplus f_2)$, and finally by those of the form $f\otimes h+h\otimes f-(f,h)$ with $f,h\in\mD$. $\mF(M,g)$ can be equipped with a natural topology induced from that on $\bigoplus_{n=0}^\infty\mD^n$. This is tantamount to the request that a sequence $f_j=\oplus_{n} f_{j,n}$ is said to converge to $f$ if and only if 
\begin{enumerate}
\item every $f_{j,n}\to f_n$ in $\mD^n$ with respect to the topology of uniform convergence of all derivatives on a fixed compact set,
\item it exists an $N\in\bN$ such that $f_{j,n}$ vanishes for every $n>N$ and for every $j$.
\end{enumerate}
In the forthcoming discussions it will sometimes also be possible to use a weaker version of $\cF(M,g)$ which is defined in the same way as $\cF(M,g)$, but without including the Dirac equations in the construction of the ideal $\cI$; 
we shall 
refer to this case as the {\em off-shell formalism}.

As a subsequent natural step, one can realise that the sesquilinear form $(\cdot,\cdot)$ can be promoted to a genuine non-degenerate scalar product on the coset space $\mD/\ker S\oplus S_*$, which, in turn, can be completed to a Hilbert space $\cH$ with respect to the said scalar product. As a by-product, this entails the possibility to extend $\cF(M,g)$ to a $C^*-$algebra $\gF(M,g)$ representing the elements as bounded operators on $\cH$ itself. Following \cite{Araki70}, we shall refer to this scenario as the assignment of the $C^*-$algebra of Dirac fields $\gF(M,g)$ to the pair $(\cH,\Ga)$. 

We have explained how to construct the local algebra in the case of Dirac fields; however, from a physical point of view, observables are required to commute at spacelike separations and the full $\cF(M,g)$ does not fulfil such requirement. As a first step towards a definition of the algebra of local observables of a Dirac field, we can restrict our attention to 
$$ \mF_{even}(M,g)\doteq\text{even subalgebra of } \mF(M,g),$$
which, {\it e.g.}, can be defined as the subalgebra invariant under $B(f)\mapsto -B(f)$ \cite{Dimock, Hollands, Sanders}. The reason to choose such a subalgebra stems from the fact that any two elements of $\mF_{even}(M,g)$ indeed commute for spacelike separations:

\proposizione\label{obs}{Let $A_i$, $i\in\{1,2\}$ be two elements of $\mF_{even}(M,g)$ which arise as finite linear combinations of smeared $B(f)$ generators $$A_i\doteq\sum_n B(f^i_{n,1})\cdots B(f^i_{n,2k_n})$$ such that 
$$\bigcup_{n,j}\text{supp }f^1_{n,j}\quad\text{and}\quad\bigcup_{n,j}\text{supp }f^2_{n,j}$$ are spacelike separated. Then
$$
[A_1,A_2]=0\;.
$$
}
\proof{
The proof descends out of two key observations: on the one hand we know the following relation between the commutator and anticommutator of four operators $A, B, C$ and $D$:
\beq\label{commut}
[AB,CD]=A\{B,C\}D-AC\{B,D\}-C\{A,D\}B+\{A,C\}DB\;.
\eeq
On the other hand we know that, given two field algebra elements $B(f)$ and $B(g)$ with the support of $f$ and $g$ spacelike separated, condition iv) in definition \ref{fieldal} together with the support properties of the causal propagator, proved in theorem \ref{fund}, entail that
$$B(f)^*B(h)+B(h)B(f)^*=0.$$
To conclude the proof, one needs to notice that, since only products of an even number of fields appear, the properties of the commutator allow to reduce $[A_1,A_2]$ to a linear combination of commutators, all of the form \eqref{commut} with $AB$ and $CD$ of the form $B(f^1_{n,j_1})B(f^1_{n,j_2})$ and $B(f^2_{n,j_1})B(f^2_{n,j_2})$ respectively. This operation together with the requirement on the supports of the test sections defining $A_1$ and $A_2$ concludes the proof.}

With the restriction to $\mF_{even}(M,g)$ we have been able to assure local commutativity. This criterion is, however, not sufficient to extract the observable elements out of $\cF(M,g)$ and $\mF_{even}(M,g)$ is thus still too large to be a candidate for the algebra of local observables. To obtain such a good candidate, we have to take only so-called ``gauge invariant'' elements of $\mF_{even}(M,g)$ into account, {\it cf.}, {\it e.g.}, \cite{Araki70, Dimock, Hollands, Sanders}, and we denote the resulting subalgebra with $\cA(M,g)$. Particularly, if we consider any $A\doteq\sum_n B(f_{n,1})\cdots B(f_{n,2k_n})$ in $\mF_{even}(M,g)$, it is lying in $\cA(M,g)$ if and only if it is invariant under the action of any $S\in Spin_0(3,1)$; such an action is defined by a straightforward extension of the known one on $DM$ and on $D^*M$, first to $DM\oplus D^*M$ and subsequently to arbitrary outer tensor products of the latter, such that we have a well-defined action on the test sections $f_{n,1}\otimes \cdots \otimes f_{n,2k_n}$ determining $A$. 
This definition of an observable is compatible with the product of algebra elements, and thus defines a subalgebra of $\mF_{even}(M,g)$ in a well-defined way. In the rest of the paper we shall always work with $\mF_{even}(M,g)$, though all our results can be applied to $\mA(M,g)$ (and to the topological closures of the mentioned algebras) as well.

It is remarkable that, in order to get to the definition of the various algebras introduced above, once a particular representation of the Clifford algebra has been chosen, the only other necessary datum is the geometry of the underlying manifold. This can be understood realising that, beside the Dirac bundles $DM$ and $D^*M$ themselves, the overall analysis relies on the causal propagators $S$ and $S_*$, which are unique in a globally hyperbolic spacetime with spin structure. This apparently innocuous observation will play an important role in identifying  the quantisation of the Dirac field as a particular locally covariant quantum field theory, as we will explain in the next subsection. 

\subsection{Locality and general covariance}\label{funct}

In order to establish a connection between the previous discussion and the modern interpretation of quantum field theory over curved backgrounds, it is mandatory to address the question whether the axioms of a locally covariant theory, as proposed by Brunetti, Fredenhagen, and Verch in \cite{BFV} are fulfilled for the above displayed algebraic quantisation of Dirac fields, and an affirmative answer has indeed been given in \cite{Sanders}. We shall not dwell on a recapitulation of the precise definition of all the needed tools, {\it e.g.}, the involved categorical notions here: instead, we choose to provide a short overview and we refer an interested reader to \cite{BFV, BrFe09, Sanders} for further details. That said, per direct inspection of the previous analysis, we can infer that the following axioms of a locally covariant theory are satisfied:
\begin{enumerate}
\item It is possible to associate to every globally hyperbolic spacetime $(M, g)$ with spin structure $(SM,\rho)$ the corresponding $*-$algebra $\mF(M,g)$ of fields in a unique way, once a global representation of the Clifford algebra has been chosen.
\item To every map $\chi$ which is an isometric and orientation preserving embedding of $(M_1,g_1)$ into $(M_2,g_2)$ and at the same time maps $(SM_1,\rho_1)$ to $(SM_2,\rho_2)$ in a coherent and equivariant way (cf. Definition 2.3.1 of \cite{Sanders}), one can associate an injective, unit-preserving $*$-homomorphism $\alpha_\chi$ between the corresponding fields algebras $\mF(M_1,g_1)$ and $\mF(M_2,g_2)$.
\item Let us choose two maps as above, namely, $\chi_1:(M_1, g_1, SM_1, \rho_1)\to (M_2, g_2, SM_2, \rho_2)$ and $\chi_2:(M_2, g_2, SM_2, \rho_2)\to (M_3, g_3, SM_3, \rho_3)$; then the following composition law is satisfied for the corresponding algebra morphisms 
$$
\alpha_{\chi_1\circ \chi_2 } = \alpha_{\chi_1} \circ \alpha_{\chi_2 } \;.
$$
\end{enumerate}

Let us note that the above axioms are also fulfilled in the off-shell formalism, {\it i.e.}, for Dirac spinor fields not subject to the Dirac equations. We can, furthermore, add another two axioms in special cases: on the one hand, if we restrict the construction to $\cF_{even}(M,g)$, the axiom of {\it Einstein causality} is fulfilled on account of proposition \ref{obs}:

\begin{itemize}
\item[4.] Consider two globally hyperbolic spacetimes with spin structure $(M_1,g_1,SM_1,\rho_1)$ and $(M_2,g_2,SM_2,\rho_2)$ together with $\chi_1$ and $\chi_2$ respectively, two embeddings into a third spacetime with spin structure $(M_3,g_3,SM_3, \rho_3)$ of the aforementioned kind. Under the assumption that $\chi_1(M_1)$ and $\chi_2(M_2)$ are spacelike separated in $M_3$, it holds that 
for every $A_1\in\mF_{even}(M_1,g_1)$ and $A_2\in\mF_{even}(M_2,g_2)$, $[\alp_{\chi_1}(A_1), \alp_{\chi_2}(A_2)]=0$.
\end{itemize}

At the same time, in the on-shell formalism - though not slavishly for the extended algebra we shall later introduce - the {\it Time slice axiom} ({\it cf.}, proposition 4.2.22 of \cite{Sanders}) holds:

\begin{itemize}
\item[5.] Let $\chi:M_1\to M_2$ be a map between two globally hyperbolic spacetimes with spin structure with the properties already discussed. If a Cauchy surface of $M_2$ is contained in $\chi(M_1)$, then $\alpha_\chi$ is an isomorphism. 
\end{itemize}

These axioms state properties of the full field algebras, but one can refine these statements and identify fields with a special behaviour under the maps $\chi$ and $\alp_\chi$, the so-called {\it locally covariant fields} \cite{BFV, HW01}. In fact, as discussed by Sanders in \cite{Sanders}, the field $B(\cdot)$, and, thus, also the single fields $\psi(\cdot)$, $\psi^\dagger(\cdot)$, are locally covariant fields. This entails that $B(\cdot)$ can be understood as family of continuous maps, indexed by spacetimes with spin structure $M$\footnote{We omit the other data determining a spacetime with spin structure in the remainder of this paragraph in favour of notational simplicity.}, $$B_M: \cD(M) \to \cF(M,g),$$ such that, given two spacetimes with spin structure $M_1$ and $M_2$ and a map $\chi: M_1\to M_2$ with the properties discussed in axiom 1., one gets the same result by either building a quantum field out of a test section $f_{M_1}\in \cD(M_1)$ and then mapping this field to $\cF(M_2,g_2)$ via $\alp_\chi$ or by mapping the test section $f_{M_1}$ to $f_{M_2}\doteq \chi_*(f_{M_1}) \in\cD(M_2)$ via the push-forward $\chi_*$ of $\chi$ and then building $B_{M_2}(f_{M_2})$ out of it. On the level of maps, we thus have: $$\alp_\chi\circ B_{M_1}=B_{M_2}\circ \chi_*.$$ Similarly, one can identify certain observable composite fields as locally covariant quantum fields via specific choices of test section spaces, and, furthermore, the Wick polynomials we shall discuss later fit into the same framework as well.

\subsection{Spinors and Hadamard states}\label{SHS}

The algebra $\mA(M,g)\subset \cF_{even}(M,g)$, is, at this point of our discussion, the best candidate to play the role of an algebra of observables for a free Dirac field theory. Unfortunately, this status is far from being satisfactory because objects such as all the Dirac bispinors, the current in particular, and the (components of the) stress-energy tensor are not contained in $\mA(M,g)$ or $\mF_{even}(M,g)$. Since we want to consider these as genuine observables, the best option is to solve this problem along the same lines employed in the scalar case, namely, we shall suitably enlarge $\mF_{even}(M,g)$ to include all the wanted elements. Although reasonable and, to a certain extent natural, such idea comes with a price to pay, {\it i.e.}, not all the well-behaved algebraic states for $\mF_{even}(M,g)$ are admissible for the extended algebra; in fact, we have to select only those with the suitable ultraviolet behaviour already possessed by the Minkowskian vacuum state.

This is indeed not a novel problem and it has been tackled by several authors; the underlying philosophy is to characterise the admissible states imposing suitable physical conditions, such as finiteness of quantum fluctuations, and thus the possibility to employ these states to define a sensible expected stress-energy tensor \cite{Wald2}. The translation of these ideas in a mathematical language leads to the notion of {\it Hadamard states}, which we shall now discuss in our framework. The available literature is immense and we point a reader interested in further details to \cite{Kay, Radzikowski, Radzikowski2, Koehler} for scalar fields or to \cite{Hollands, Kratzert, SahlmannVerch} for a discussion related to spinors. 

As a first step and as main topic of this section, we shall proceed introducing the notion of Hadamard states for the whole $\mF(M,g)$ and only later we shall restrict them to $\mF_{even}(M,g)$. The already anticipated enlargement of the algebra to include all interesting observables of the free field will then be the core of a subsequent discussion. 

That said, henceforth, we shall consider a state $\om$ to be a continuous, positive, and normed linear functional on $\mF(M,g)$, such that $$\om(A^*A)\ge 0\quad \forall A\in \mF(M,g),\quad \om({\bf 1})=1;$$ since this algebra is generated, according to definition \ref{fieldal}, by the abstract elements $B(f)$ with $f\in\mD$, every said state is uniquely determined by the set of its $n$-point functions, namely,
$$
\om_n(f_1,\dots,f_n)\doteq \om(B(f_1) \dots B(f_n))
$$
where, due to the required properties a state, each $\om_n$ is a distribution on $\cD^n$. The bridge between the algebraic formulation of quantum field theory employed in this work and its usual Hilbert space description is in the non-trivial direction provided by the {\it Gelfand-Naimark-Segal (GNS) construction} (c.f., {\it e.g.}, \cite{Haag}) which yields a representation of an algebraic state and a field algebra in terms of a Hilbert space state and operator valued distributions on the same Hilbert space respectively. Among all possible algebraic states, a distinguished role is played by the so-called {\it quasi-free} ones, whose $n$-point functions can be determined fully out of $\om_2$. We shall focus on these, and, following \cite{Araki70}, we recall: 

\definizione\label{quasifree}{A state $\omega:\mF(M,g)\to\bC$ is called {\bf quasi-free} if, given any set of $f_i\in\mD$ with $i \in\{1,\cdots,n\}$, $\omega\left(B({f_1})\cdots B({f_{n}})\right)$ vanishes for odd $n$ while for even $n$ it holds
$$
\omega\left(B({f_1})\cdots B({f_{n}})\right)=
\sum\limits_{\pi_n\in S'_n}(-1)^{|\pi_n|}\prod\limits_{i=1}^{n/2}\omega_2\left(f_{\pi_n(2i-1)},f_{\pi_n(2i)}\right).
$$
Here, $S'_n$ denotes the set of ordered permutations of $n$ elements, namely, the following two conditions are satisfied for $\pi_n\in S'_n$:
$$
\pi_n(2i-1)<\pi_n(2i), \qquad 1\leq i\leq n/2, 
$$ 
$$
\pi_n(2i-1)<\pi_n(2i+1). \qquad 1\leq i <  n/2\;.
$$}

\noindent Even though it is in principle possible to state the Hadamard property for general states \cite{Strohmaier, Sanders}, we will restrict our discussion to quasi-free ones and can thus state everything on the level of two-point functions. In this context and on the level of single Dirac fields, two distinguished distributions appear: 
\beq\label{2pf}
\omega^+(f,h)\doteq \omega\left(\psi(h)\psi^\dagger(f)\right)
\quad\textrm{and}\quad
\omega^-(f,h)\doteq \omega\left(\psi^\dagger(f)\psi(h)\right),
\eeq
where $f\in \mD(DM)$ whereas $h\in\mD(D^*M)$ and where both $\psi^\dagger(f)$ and $\psi(h)$ are particular elements of $\mF(M,g)$ as explained in subsection \ref{laqft}. Hence, it turns out that both $\omega^+$ and $\omega^-$ can be understood as distributions on $\mD(DM \boxtimes  D^*M)$. 
 
We can now introduce the notion of Hadamard state and, as in the scalar case, it is remarkable and useful that this concept can be illuminated in two equivalent ways. The first one has recourse to the notion of {\it wavefront sets} \cite{Duistermaat, Hormander}, a concept which enables a refined formulation of a singularity structure of a distribution, and, to this avail, one should take into account that an {\it a priori} obstacle lies in the nature of the vector-valued distributions appearing in the context of Dirac fields. Particularly, since wavefront sets are more familiar in the context of scalar distributions, we need to specify how they can be defined for distributions with values in higher-dimensional spaces. To achieve this, it appears to be natural to define the wavefront set of a vector valued distribution as the union of the wavefront sets of the coefficients with respect to a (possibly local) basis-expansion and indeed this turns out to be an invariant concept due to the properties of scalar wavefront sets \cite{Dencker, Kratzert, SahlmannVerch}. Specifically, we can define the wavefront sets of $\om^\pm(x,y)={\om^\pm}_A^{\;\;\;B'}(x,y)\,E^A(x)\otimes E_{B'}(y)$ as $$WF(\om^\pm)\doteq \bigcup\limits_{A=1}^4\bigcup\limits_{B'=1}^4 {\om^\pm}_A^{\;\;\;B'}(x,y).$$
By defining wavefront sets in this way, we certainly loose information on the most singular ``directions'' of a vector-valued distribution. This information can be encoded in so-called {\it polarised wavefront sets}, as introduced in  \cite{Dencker} and applied in \cite{Kratzert, Hollands}. Though of high mathematical interest, such concept is of no use in our
approach and we feel safe not to dwell into it since we would end up providing only shallow ideas. That said, let us state the first possible access to Hadamard states \cite{Radzikowski, Kratzert, Hollands, SahlmannVerch}:

\definizione\label{globHad}{A quasi-free state $\om$ satisfies the {\bf microlocal spectral condition} ($\mu$SC) and is thus called a {\bf Hadamard state} if only if
\begin{gather*}
WF(\omega_2)=\left\{(x,y,k_x,-k_y)\in T^* M^{\boxtimes 2}\setminus 0,\;|\;
(x,k_x)\sim(y,k_y),\quad k_x\triangleright 0\right\}.
\end{gather*}
Here, $(x,k_x)\sim(y,k_y)$ implies that it exists a null geodesic $\gamma$ connecting $x$ to $y$ such
that $k_x$ is coparallel and cotangent to $\gamma$ at $x$ and $k_y$ is the parallel transport of $k_x$ from $x$ to $y$ along $\ga$. Finally, $k_x\triangleright 0$ means that the covector $k_x$ is future-directed.
}

\remarks{If a quasi-free state $\om$ fulfils the $\mu$SC, then $\om^\pm$ possess the following wavefront sets:
$$
WF(\omega^\pm)=\left\{(x,y,k_x,-k_y)\in T^*M^{\boxtimes 2} \setminus 0,\;|\;
(x,k_x)\sim(y,k_y),\quad k_x \stackrel{\triangleleft}{_\triangleright} 0\right\} \;,
$$
where $k_x\triangleleft 0$ states that $k_x$ is past-directed.\\\indent
An even stronger relation between the two distributions $\om^\pm$ arises if we employ the anticommutation relation since it entails that 
$$
\omega^+(f,h) + \omega^-(f,h)=-i\langle h,Sf\rangle\;.
$$
\indent By contrast, the distributions $\om(\psi(h_1)\psi(h_2))$ and $\om(\psi^\dagger(f_1)\psi^\dagger(f_2))$, which, together with $\om^\pm$ determine $\om_2$, have smooth integral kernels. For $\om(\psi(h_1)\psi(h_2))$, this can be proved employing a symmetry argument already used in a similar way in \cite{Radzikowski}: due to the anticommutation relations, we have
$$\om\left(\psi(h_1)\psi(h_2)\right) = -\om\left(\psi(h_2) \psi(h_1)\right).$$
Hence, if $(x,y,k_x,k_y)$ is an element of the wavefront set of the distribution on the right hand side of the previous equation, then $(y,x,k_y,k_x)$ must lie in the wavefront set of the other one. At the same time, on account of the $\mu$SC, we know that $WF\left(\om(\psi(x)\psi(y))\right)$ is not invariant under the exchange of coordinates. This entails that $WF\left(\om(\psi(x)\psi(y))\right)=\emptyset$, hence, $\om(\psi(h_1)\psi(h_2))$, and analogously $\om(\psi^\dagger(f_1)\psi^\dagger(f_2))$, possesses a smooth integral kernel.}

Although highly elegant from a mathematical point of view and thus very helpful in abstract proofs, the microlocal definition of a Hadamard state is
neither the first one introduced chronologically nor the easiest one to cope with on the level of explicit calculations. In fact, as already promised at the beginning of this section, there is a different, more explicit definition of a Hadamard state via the so-called {\it Hadamard form}. For scalar fields, this has been rigorously introduced in \cite{Kay}, while for Dirac fields a similar concept has been proposed by \cite{Koehler, Verch}. To introduce it, we need the notion of a {\it convex normal neighbourhood}, which is an open subset $\cO$ of $M$ such that any two points $x, y \in \cO$ can be connected by a unique geodesic. On any convex normal neighbourhood, we can introduce the smooth halved squared geodesic distance $\si(x,y)$, and, finally, formulate the following definition:

\definizione{\label{hadamardform} A quasi-free state $\om$ is said to be of the {\bf Hadamard form} if and only if in any convex normal neighbourhood the distributions kernels of $\om^\pm$ can be written as 
$$\om^\pm(x,y)=\frac{1}{8\pi^2}\pm D'_y\left(H^\pm(x,y)+W(x,y)\right),$$
where the index $\cdot_y$ stresses that the dual Dirac operator $D'_y$ acts on the $y$-variable, and the singular Hadamard distribution kernels $H^\pm$ can be specified as
\beq\label{singstr}
H^\pm(x,y)=\frac{U(x,y)}{\sigma_{\pm\epsilon}(x,y)}+V(x,y)\ln\frac{\sigma_{
\pm\epsilon(x,y)}}{\lambda^2}.
\eeq
Here, $U$, $V$, as well as $W$ are smooth bispinors and $V$ and $W$ can be expanded in powers of $\si$, {\it viz.},
$$V(x,y)\doteq\sum\limits_{n=0}^\infty V_n(x,y)\sigma(x,y)^{n},\quad W(x,y)\doteq\sum\limits_{n=0}^\infty W_n(x,y)\sigma(x,y)^{n},$$ 
where $\lambda$ is a reference length, and $\sigma_{\pm\epsilon}(x,y)\doteq \sigma(x,y)\pm 2i\epsilon\left(T(x)-T(y)\right)+\epsilon^2$ with $\epsilon>0$. In the above formula, $T$ is a time function, such that $\nabla T$ is timelike and future pointing on the full spacetime $(M,g)$.\\\indent
We furthermore require $H^\pm$ to be bisolutions of the spinorial Klein-Gordon equations up to smooth terms, {\it i.e.}, \beq\label{eqationscalar}P_xH^\pm(x,y)\in \cE(DM\otimes D^*M),\quad P_yH^\pm(x,y)\in \cE(DM\otimes D^*M)\eeq and demand that their difference is specified by the fundamental solution of $P$, {\it viz.},$$H^+(f,g)-H^-(f,g)=i\langle g,Ef\rangle,$$ where $f\in \cD(DM)$ and $g\in \cD(D^*M)$.}

\remarks{The existence of a time function $T$ is guaranteed on any globally hyperbolic manifold \cite{Bernal, Bernal2}.\\
\indent Furthermore, a completely satisfactory definition of the Hadamard form requires some more work to rule out spacelike singularities, to circumvent convergence problems of the series $V$ and $W$, which are only asymptotic, and, finally, to assure that the definition does not depend neither on a special choice of the temporal function $T$ nor on the employed convex normal neighbourhood. For further details and discussions of these aspects and the existence of states of the Hadamard form we refer an interested reader to \cite{Fulling, Kay, Koehler, Verch, SahlmannVerch}.}

To determine the so-called {\it Hadamard coefficients} $U$, $V$, and $W$, one has to exploit the equations \nref{eqationscalar}. At this point, we would like to stress a slight conceptual difference between Dirac spinors and scalar fields: in the case of scalar fields, the two-point function fulfils the Klein-Gordon equation in both entries, and this property is thus inherited by its singular Hadamard kernel up to smooth terms. Contrariwise, in case of Dirac spinors, \nref{eqationscalar} does \underline{not} follow straightforwardly from the fact that the two-point functions $\om^\pm$ fulfil the Dirac equations. In more specific terms, if we recall the definition of $\om^\pm$ \eqref{2pf}, we know that they fulfil 
$$
D'_x\om^\pm(x,y)=D_y\om^\pm(x,y)=0.
$$ 
Consequently, 
\begin{gather*}D'_xD'_y\left(H^\pm(x,y)+W(x,y)\right)=0, \\ D_yD'_y\left(H^\pm(x,y)+W(x,y)\right)=-P_y\left(H^\pm(x,y)+W(x,y)\right)=0,\end{gather*} 
and, thus, both $P_yH^\pm$ and $D'_xD'_yH^\pm$ are smooth. The smoothness of $P_xH^\pm$ does however, not follow automatically from these considerations, but has to be required or proven in a way similar to the one displayed in lemma 5.4. of \cite{SahlmannVerch}. 

We shall explicitly discuss the computation of $U$, $V$ and $W$ in appendix \ref{Hadcoeff}. To this avail, the following proposition will prove to be very helpful:
 
\proposizione{\label{extrasmooth}
Let $H^\pm(x,y)$ be the Hadamard distribution kernels of a state introduced in definition \ref{hadamardform}. Then $(D_x-D'_y)H^\pm(x,y)=(D_y-D'_x)H^\pm(x,y)$ are smooth.
}

\proof{The overall strategy calls for combining a deformation argument as devised in the appendix C of \cite{Fulling} together with the so-called theorem of propagation of singularities ({\it cf.} theorem 6.1.1 in \cite{Duistermaat}).\\\indent 
That said, let us proceed in logical sequential steps and consider any Cauchy surface $\Si\hookrightarrow (M,g_{\mu\nu})$ of the spacetime we are interested in and let us choose an open neighbourhood of $\Si$, say $\mO_\Si$, such that it is a causal normal neighbourhood of $\Si$, {\it i.e.}, $\Si$ is a Cauchy surface for $\mO_\Si$ and for each $p,q\in\mO_\Si$ such that $p\in J^+(q)$, it exists a convex normal neighbourhood containing $J^-(p)\cap J^+(q)$. The existence of such sets in a globally hyperbolic spacetime and for any Cauchy
surface $\Si$ was first proved in \cite{Kay}.\\\indent
The above mentioned deformation argument grants us that it is possible to construct an isometric, orientation and time-orientation-preserving embedding, say $\chi$, of $\mO_\Si$ in causal normal neighbourhood $\mO_{\Si'}$ of a Cauchy surface $\Si'$ of a second globally hyperbolic spacetime $M^\prime$. Furthermore, one can engineer $M^\prime$ in such a way that, in the past of $\chi(\mO_\Si)$, it exists another Cauchy surface $\Si''$ with a neighbourhood $\mO_{\Si''}$ which contains the image of a suitable neighbourhood of a Cauchy surface $\Si'''$ in Minkowski spacetime under an isometric, orientation preserving, embedding $\widetilde\chi$, and it is straightforward to extend $\chi$ and $\widetilde\chi$ in such a way that they respect the spin structures.\\\indent
Since $H^\pm$ on $\mO_\Si\times\mO_\Si$ are constructed only out of the local geometric data via \nref{eqationscalar}, it is possible to build a second pair $\widetilde H^\pm$ which coincides with the push-forward under $\chi$ of $H^\pm$ in $\chi(\mO_\Si)\times\chi(\mO_\Si)$. Furthermore, due to the propagation of the Hadamard form as proved in \cite{Fulling, SahlmannVerch}, $\widetilde H^\pm$ are of the Hadamard form in $\mO_{\Si''}\times\mO_{\Si''}$ as well, and their pull-back to $\mO_{\Si'''}\times\mO_{\Si'''}$ thus coincide with the Hadamard distribution kernels in Minkowski spacetime.\\\indent
Let us now consider $$u^\pm\doteq(D_x-D'_y)H^\pm$$ and proceed to prove that these distributions have empty wavefront set. According to the above discussions, we can push-forward $u^\pm$ to $M^\prime$ and subsequently pull them back to the neighbourhood $\mO_{\Si'''}$ in Minkowski space in a well-defined way. In the Minkowskian region, the pushed-forward and then pulled-back versions of $u^\pm$ are identically vanishing and thus have empty wavefront set, since the flat spacetime Hadamard kernel only depends on $x-y$ due to translational invariance. To access the wavefront set of $u^\pm$ in the original spacetime region under interest, let us note that these distributions satisfy $P_xP_y u^\pm =0$ up to smooth terms and the same holds for their mentioned push-forwards and pull-backs, where the operator $P_xP_y$ is properly supported, of real
principal type, and homogeneous of degree $2$ since it is the tensor product of two second order hyperbolic
differential operators. From this it follows due to the contravariant transformation properties of wavefront
sets under diffeomorphisms and the propagation of singularities theorem (see \cite{SahlmannVerch} or chapter
8 of \cite{Hormander}) that the wavefront set of $u^\pm$ on $\mO_\Si\times \mO_\Si$ can only contain elements
of the form \beq\label{half}(x,y,k_x,0)\quad\text{or}\quad (x,y,0,k_y).\eeq Following a line of argument
employed in the proof of theorem 5.8 in \cite{SahlmannVerch}, we can infer that $WF(u^\pm)=\emptyset$ in the following way: since $u^\pm$ are constructed as Dirac derivatives of $H^\pm$ and properly supported partial differential operators like $D$ and $D'$ do not increase the wavefront set, we know that $WF(u^\pm)\subset WF(H^\pm)$. If we furthermore recall that $H^\pm$ specify the singular parts of $\om^\pm$ and that these kernels have the "antisymmetric" wavefront set displayed in definition \ref{globHad}, it follows that $WF(u^\pm)$ can not even contain elements of the form \nref{half} end are thus empty.}

As a result of the procedures described in appendix \ref{Hadcoeff}, $U$ and $V$ turn out to depend only on the local geometry and the mass $m$, while the state dependence of $\om^\pm$ is encoded in $W$. This ``universality'' of the singularity structure of states of the Hadamard form allows for a locally covariant definition of normal ordering, as we will see in the next subsection. To this avail, it will be useful to compose the Hadamard distributions to a single object living on $\cD\otimes \cD$, {\it viz.},
\begin{align}\label{bihadamard}
\gH(f_1\oplus f_2, h_1\oplus h_2)&\doteq \left(D'_yH^+\right)(h_1,f_2)-\left(D'_yH^-\right)(f_1,h_2)\\&=H^+(h_1,Df_2)-H^-(f_1,Dh_2)\notag,
\end{align}
where $f_1\oplus f_2$, $h_1\oplus h_2$ $\in\cD$. Before we start working with Hadamard states, let us state the already anticipated and fruitful equivalence of the Hadamard form and the $\mu$SC, which is a result due to \cite{Kratzert, Hollands, SahlmannVerch}:

\teorema{
Let us consider a state $\om$ on $\cF(M,g)$ with two-point function $\om_2$. This satisfies the microlocal spectral condition, if and only if the distribution on $\cD \otimes \cD$ defined by
\beq\label{hadequiv}
f\otimes h\;\mapsto\; \om_2(f,h)-\gH(f,h),
\eeq
has a smooth integral kernel, and, thus, $\om^\pm$ are of Hadamard form.
}

To conclude this section, we would like to mention a most notable property of Hadamard states: despite the well-known problem to fix a unique vacuum state for a quantum field theory on a generic spacetime, Hadamard states turn out to be ``almost'' unique since they are all locally quasi-equivalent \cite{VerchQ, Verch, DaHo}. This implies that locally the density matrix states on the Hilbert spaces obtained from Hadamard states via the GNS construction are all equal, and, in more physical terms, that any two Hadamard states have a finite energy density with respect to one another. The latter statement is of course related to the expected stress-energy tensor, the main topic of the last section of this work.

\subsection{On the notion of Wick polynomials}\label{wicks}

In the development of quantum field theory, a well-known obstruction arises whenever we
consider the product of two fields, which, being distributions, cannot be safely multiplied unless special
conditions are met. Since, as already anticipated, our ultimate goal is to enlarge the algebras under consideration to include observables such as the stress-energy tensor for Dirac fields, we are lead to tackle this problem. Like in the case of scalar fields, this results in the introduction of Wick polynomials and in the following we shall try to adapt an approach similar to the one discussed in the work of Brunetti, Duetsch, and Fredenhagen \cite{BDF} which in turn is related to further earlier works \cite{BF00, HW01, HW02}.

Unsurprisingly, in our scenario, there are differences to the above mentioned works due to the vectorial nature of our fields and their anticommutativity. This necessitates a treatment of Wick polynomials of Dirac fields on curved spacetimes on its own and we will thus develop them in this subsection since they have not been treated in the literature in the past. 

As already anticipated, upon enlargement of the field algebra to include Wick polynomials we have to restrict our state space to Hadamard states, which seems not to be a real loss since these are already distinguished and presumably the only ``physical'' ones for the algebras discussed in subsection \ref{laqft}.

As a starting point to define the extended algebra of fields, it will be more convenient not to start directly from $\mF(M,g)$ or its subalgebras, though we shall consider the set
$$
\cC(M,g)\doteq\bigoplus_{n=0}^\infty\mD^n_A, 
$$
where the subscript $A$ indicates that, for $n>0$, one takes into account only antisymmetric elements, while $\mD^0_A=\mD^0=\bC$. Notice that it is required that a generic element $F\in\cC(M,g)$ is unambiguously determined by a {\em finite} sequence $\{ F^{(n)}\}$ of antisymmetric elements lying in $\cD^n$. This entails that, in a basis expansion with respect to $\cE^\Th(x)\doteq E_A(x)\oplus E^B(x)$, each element $F^{(n)}=F^{(n)}_{\Theta_1\cdots\Theta_n}\cE^{\Theta_1}\otimes \cdots \otimes \cE^{\Theta_n}$ has antisymmetric coefficients, {\it viz.}, 
\begin{align}\label{antisym}&
F^{(n)}_{\Theta_1,\dots,\Theta_k,\Theta_{k+1},\dots,\Theta_n}(x_1,\dots, x_k,x_{k+1},\dots ,x_n)\notag\\ =&
-F^{(n)}_{\Theta_1,\dots,\Theta_{k+1},\Theta_k,\dots,\Theta_n}(x_1,\dots, x_{k+1},x_k,\dots ,x_n) \quad \forall 1\leq k \le n.
\end{align}

The set $\cC(M,g)$ can be promoted to an algebra with respect to the following product which we shall henceforth indicate as $\cdot_A$; let $F\doteq\{F^{(n)}\}$ and $G\doteq\{G^{(n)}\}$ be two generic elements in $\cC(M,g)$, then  
\beq\label{asymprod}
(F \cdot_A G)^{(n)}\doteq \sum_{p+q=n} \bA \at F^{(p)} \otimes G^{(q)} \ct ,
\eeq
where $\bA$ is the operator of total antisymmetrisation such that $F \cdot_A G$ is indeed an element of $\cC(M,g)$. 
Specifically, $\bA$ leaves $\cD^0$ invariant, while, for an arbitrary $F^{(n)}\doteq f_1 \otimes\dots  \otimes f_n$ with $f_i\in \cD$ and $n>0$, the antisymmetrisation reads
$$
\bA(f_1 \otimes\dots  \otimes f_n)=\frac{1}{n!}\sum_{\pi_n\in S_n} (-1)^{|\pi_n|} f_{\pi_n(1)} \otimes \dots \otimes f_{\pi_n(n)},
$$
where the sum is taken over all permutations\footnote{Of course not all permutations employed are necessary, since in \eqref{asymprod} the constituents will already be antisymmetric. The antisymmetrisation as defined here, however, is still valid and it constitutes the easiest way to write it without unnecessarily getting lost in combinatorics.} $\pi_n\in S_n$ and $\bA$ can be extended to $\cD^n$ by linearity and continuity. The algebra $(\cC(M,g), \cdot_A)$ can be interpreted as the algebra of functionals, in the sense of distributions on smooth sections, on the classical field configurations of Dirac spinors.

The standard quantisation scheme is eventually realised changing the product $\cdot_A$ into a suitable 
$\star$-product compatible with the anticommutation relations. The overall procedure, once a functional $\De:\cD^2\to\bC$ is selected, can be realized out of the map $\Ga_{\De}:\cD^n\to \cD^{n-2}$ whose action on a generic element $F^{(n)}$ of $\cD^n$ (notice, here taken without antisymmetrisation),
is required to be trivial if $n<2$, whereas, for $n\ge 2$,
\begin{align}
\Ga_{\De} F^{(n)} \doteq & \sum_{i=1}^{n-1} \sum_{j=i+1}^n \int\limits_M d\mu(x_{i})\int\limits_M  d\mu(x_{j})\, (-1)^{j-i+1} \De^{\Th_i\Th_j}(x_{i},x_{j}) F^{(n)}_{\Th_1\cdots\Th_n}(x_1,\dots,x_n) \;\times\nonumber\\
\times & \;\cE^{\Th_1}(x_1)\otimes \dots \otimes \widehat{ \cE^{\Th_i}(x_i)}\otimes   \dots \otimes\widehat {\cE^{\Th_j}(x_j)} \otimes  \dots  \otimes \cE^{\Th_n}(x_n).\label{twist}\end{align}

Here, $\De(x_i,x_j)=\De^{\Th_i\Th_j}(x_{i},x_{j})\,\cE_{\Th_i}(x_i)\otimes\cE_{\Th_j}(x_j)$, with $\cE_{\Th}(x)\doteq E^A(x)\oplus E_B(x)$, denotes the integral kernel of $\De$, whereas the symbol $\widehat{\cE^{\Th_i}(x_i)}$ indicates that $\cE^{\Th_i}(x_i)$ must be omitted.

On account of the regularity of $\cC(M,g)$, we can safely define a $\star$-product as 
\beq
F \star_S G = \bA \ \alpha_{S} \at  F  \otimes   G \ct,\label{star1}
\eeq
where $\alpha_{S}$ is defined as a formal exponential\footnote{Since $\cC(M,g)$ contains only finite sequences of test sections, the exponential series will always terminate after finitely many terms.} $$\alpha_{S}\doteq\exp \at i \frac{1}{2}\Ga_{\widetilde  S}\ct;$$
here, $\Ga_{\widetilde S}$ arises from \eqref{twist} if one inserts for $\De$ the functional $$\cD^2\ni f_1\oplus f_2 \otimes g_1 \oplus g_2\;\mapsto\; \widetilde{S}(f_1\oplus f_2,g_1\oplus g_2)\doteq-\langle f_2,Sg_1\rangle + \langle S_* f_1^\dagger, g_2^\dagger\rangle,$$ with $S$ and $S_*$ being the causal propagators constructed in theorem \ref{fund}.

\remarks{If we introduce a $*$-operation on $(\cC(M,g),\star_S)$ via the straightforward tensorialisation of $\Ga$ \eqref{bidagger}, the result is naturally isomorphic to the off-shell version of $\cF(M,g)$ with its standard product. Particularly, $B(f)\in \cF(M,g)$ corresponds to $f\in \cC(M,g)$ and the anticommutation relations on $\cF(M,g)$ correspond to $$f\star_S g + g\star_S f = i\widetilde{S}(f,g)=(\Ga f, g),$$ as follows by straightforward computation. The equations of motion can then be implemented by dividing out a suitable ideal. Since the Dirac equations will not be necessary in the following discussion, we will denote both the on-shell and off-shell algebras with $(\cC(M,g),\star_S)$ and we shall consider them as being isomorphic respectively to $\cF(M,g)$ and to its off-shell version.}

Up to now we focused on rather regular objects constructed out of $\cD$, but, alas, this does not suffice to reach our ultimate goals;  as a matter of fact, we need to consider the spaces of compactly supported distributions\footnote{Notice that elements of $\cE'$ test compactly supported sections on $D^*M\oplus DM$ and not on $DM\oplus D^*M$. This ``dual'' notation, as employed, {\it e.g.}, in \cite{Dimock}, is used to stress that $\cD\hookrightarrow \cE'$.} as well:
$$\cE^{'0}\doteq \bC,\quad \cE^{'n}\doteq\cE'\left((DM\oplus D^*M)^{\boxtimes n}\right),\quad \cE'\doteq \cE^{'1}.$$

The underlying leitmotiv is rooted in our interest in objects like $\int_M d\mu(x) f(x) \wick{\psi^\dagger(x)\psi(x)}$ which will in the subsequent discussion correspond to distributions like $f(x)\de(x,y)$, these are nothing but elements of $\cE^{'n}$ which are supported on the {\it thin diagonal}
$$Diag_n\doteq\{(x_1,\dots,x_n)\in M^n\,|\, x_1=\dots =x_n\}.$$ 
Since this amounts to potentially ill-defined operations such as taking the product of distributions at the same spacetime point, we cannot blindly extend $(\cC(M,g),\star_S)$ (and equivalently $\cF(M,g)$) to incorporate these new objects into an enlarged $*-$algebra, but we have to require some suitable regularity conditions. 

\definizione\label{extfield}{We call {\bf extended set of functionals} $\cC_{ext}(M,g)$ the set containing finite linear combinations of compactly supported distributions $F^{(n)}\in\cE^{'n}$ whose wave front set satisfies the following requirement
$$
WF(F^{(n)})\cap (M^n\times (\overline{V}_+^n\cup \overline{V}_-^n)) =\emptyset
$$
where $\overline{V}_+$ and $\overline{V}_-$ are the closure of the future and the past light cone respectively in the fibre of the cotangent bundle at each point of $M$.}

In order to adopt this definition, we have to make sure that $\cC_{ext}(M,g)$ can be made into an algebra and it is manifest that the product $\star_S$ is not up to the task since it would lead us to pointwise product of causal propagators, which is ill-defined due to their wavefront set. In order to avoid the aforementioned problem, we can replace $\star_S$ by  $\star_H$, which is nothing but \eqref{star1} with $\widetilde S$ replaced by $-2i\gH$ as defined in \eqref{bihadamard}. This new $\star$-product is equivalent to the old one when this is well-defined, being 
\begin{gather}
F\star_H G = \alpha_H \at  \alpha^{-1}_H(F)\star_S \alpha^{-1}_H(G) \ct,\label{starH}\\
\alpha_H\doteq\exp {\Ga _{H}},\notag
\end{gather}
where $\Ga_H$ is defined as in \eqref{twist} upon inserting $\gH-i/2\widetilde S$ for $\De$.
If we take into account the singular structure of $\gH$ and the criterion to multiply distributions as devised by H\"ormander ({\it cf.}, \cite{Duistermaat} or chapter 8 in \cite{Hormander}), it turns out that the problem with $\star_S$ disappears since the pointwise product of the integral kernel of $\gH$ with itself is a well defined distribution.

\remark{The outcome of the preceding discussion is the introduction of $(\cC_{ext}(M,g), \star_H)$. Recalling that $(\cC(M,g), \star_S)$ has been isomorphic to $\cF(M,g)$, we can reverse this viewpoint and just define the extended $*$-algebra $\cF_{ext}(M,g)\doteq (\cC_{ext}(M,g), \star_H)$. Similarly, restricting the possible test sections and distributions taken into account, we can define the extended algebras $\cF_{even,ext}(M,g)$ and $\cA_{ext}(M,g)$. That said, following slavishly the analysis of the scalar case in \cite{HW01, HW02}, the product \eqref{starH} rephrases {\em Wick formula} in the Dirac scenario.
}

As already announced in the prior discussion, due to the form of the wavefront set of Hadamard states, we can extend them and only them to genuine states for $\cF_{ext}(M,g)$. Particularly, this entails the standard paradigm according to which the product of two fields, say $\psi^\dagger(x)\psi(y)$, should be regularised as $$\wick{\psi^\dagger(x)\psi(y)}\;\doteq\;\psi^\dagger(x)\psi(y)+\frac{1}{8\pi^2}D'_yH^-(x,y)$$ such that, for a Hadamard state $\om$, $\om(\wick{\psi^\dagger(x)\psi(y)})=(8\pi^2)^{-1}W(x,y)$. At a level of expectation values, this can be equivalently seen as leaving $\psi^\dagger(x)\psi(y)$ unchanged while $\omega^-(x,y)$ becomes $\omega^-(x,y)+D'_yH^-(x,y)$. This somehow heuristic comment prompts the following:

\definizione{Consider a quasi-free Hadamard state $\omega$, whose $n$-point function is indicated as $\omega_n$. One can define the {\bf regularised $n$-point function} $\wick{\om_n}$ as 
\begin{gather*}
\wick{\om_n}\;\doteq\;\om_n=0\;,\qquad {\text{if $n$ is odd }}  \\
\wick{\om_n}(x_1,\dots, x_n)\;\doteq\;\sum\limits_{\pi_n\in S'_n}(-1)^{|\pi_n|}\prod\limits_{i=1}^{n/2}\left(\omega_2-\gH\right)\left(x_{\pi_n(2i-1)},x_{\pi_n(2i)}\right) \quad {\text{if $n$ is even, }}  
\end{gather*}
with $\gH$ as in \eqref{bihadamard} whereas the set of ordered permutations $S'_n\subset S_n$ is the one introduced in definition \ref{quasifree}.
}

\remark{As a straightforward consequence of the last definition, we can form expectation values of all elements in $\cF_{ext}(M,g)$.  Specifically, for any $F\doteq\{F^{(n)}\}\in \cF_{ext}(M,g)$, 
$$
\om(F)\doteq \sum_n\langle F^{(n)},\wick{\omega_b}\rangle
$$
is well defined due to the wavefront set properties of both the state and the allowed $F$. 
}

At this stage we need to point out that there are still some ambiguities in the employed definition of $H^\pm$ and thus in the definition of both $\gH$ and $\wick{\om_n}$; indeed, the reference length $\la$ necessary to construct $H^\pm$ according to definition \ref{hadamardform} is in principle undetermined. This fact does, however, not hamper our analysis since different choices of $\la$ and thus of $\gH$ lead to isomorphic algebras. 

\lemma{Suppose we choose two different $\gH$, say $\gH_1$ and $\gH_2$, to construct the extended algebra $(\cC_{ext}(M,g),\star_H)$. Then the two resulting algebras $(\cC_{ext}(M,g),\star_{H_1})$ and $(\cC_{ext}(M,g),\star_{H_2})$ are isomorphic.
}

\proof{Due to the properties of the Hadamard distributions $H^\pm$, one knows that the difference $d\doteq \gH_2-\gH_1$ has a smooth 
antisymmetric integral kernel. The two products $\star_{H_1}$ and $\star_{H_2}$ are related by a deformation. They are thus equivalent and the operator intertwining them can be realised as
$$
\alpha_d\doteq\exp\left({\Ga_d}\right),
$$
where $\Ga$ is taken as in \nref{twist} with $d$ being inserted in place of $\De$. Particularly,
$$
F\star_{H_2} G =\alpha_d \at  \alpha^{-1}_d F \star_{H_1} \alpha^{-1}_d G \ct
$$
which is well-defined and holds true since $\alpha_{H_2} \circ \alpha_{H_1}^{-1} = \alpha_d$ and $d$ has a smooth integral kernel.}

To finish the preparations for the final section of this work, we have to address a last issue. At the moment we are falling one step short from our ultimate goal since, to study the regularisation of the stress-energy tensor, one has to understand the treatment of differentiated fields. Hence, a small  addendum to the above analysis is needed and we shall follow the procedure employed for scalar fields in \cite{Mo03}, though adapted to our language. Thus, let us take a differential operator $K$ on $\mD(DM)$ of the form
$$K=a_0+\nabla_{a_1}+...\nabla^R_{a_R}, \quad R<\infty,$$
where $$\nabla^k_{a_k}\doteq a^{\mu_1...\mu_k A}_{k\quad\quad\;\;B}\nabla_{\mu_1}...\nabla_{\mu_k}$$
and $a^{\mu_1...\mu_k A}_{k\quad\quad\;\;B}$ for $k\in\{0,\cdots,R\}$ are the coefficients of an element of
$$\Gamma(\underbrace{TM\otimes\cdots\otimes TM}_{k}\otimes DM\otimes D^*M).$$
Notice that this class of differential operators encompasses both the Dirac operators and the spinorial Klein-Gordon operator which will appear in the expression of
the stress-energy tensor and of its trace. In an analogous way we can choose differential operators $K'$ on $\mD(D^*M)$ and combine them with a $K$ to operators  $K\oplus K'$ on $\mD$. If we now bear in mind definition \ref{extfield}, we realise that the extended set of fields is defined out of a condition on the wavefront set of its elements. Thus, in order to engineer any operator of the form $K\oplus K'$ into the above discussion, we just need to recall a general result on wavefront sets (cf. Chapter 8 of \cite{Hormander} or \cite{Duistermaat}) according to which a partial differential operator which is properly supported, such that it maps compactly supported distributions to compactly supported ones, does not increase the wavefront set of a distribution it is applied on. Since the differential operators we are considering are properly supported, we can readily conclude that operators of the form $K\oplus  K'$ map $\cC_{ext}(M,g)$ to itself and that the previous discussion has already encompassed the treatment of differentiated fields. One could now
prove several further properties of Wick polynomials of differentiated fields, but we will not indulge in
this task since it will play no role in the forthcoming discussion and, furthermore, the results are by all means a straightforward extension, both as concepts and as technical proofs, of those discussed in \cite{Mo03} for the scalar case.

Before proceeding with the discussion of the stress-energy tensor, let us finally remark on how the extended algebra $\cF_{ext}(M,g)$ fits into the framework discussed in subsection \ref{funct}. Without going much into details we would like to point out that any coinciding point limits of smooth objects constructed out of the Hadamard distributions are locally covariant, since the construction of $H^\pm$ depends only on the mass and the local curvature. As a result, all elements of $\cF_{ext}(M,g)$ which correspond to distributions with support on the thin diagonal are locally covariant fields.

\section{The stress-energy tensor of Dirac fields}
The aim of the section is to focus on the structure of the stress-energy tensor and to study its quantum properties. Particularly, we shall display that it is possible to introduce an improved tensor which is conserved also at a quantum level, although its trace acquires new and classically unexpected terms of geometric origin which lie at the heart of the so-called {\it trace anomaly}.

\subsection{The classical stress-energy tensor} We start our analysis by revising the form and the properties of the stress-energy
tensor for Dirac spinors in a classical framework. The Dirac equations \eqref{Dirac} can be realised as the extremal of the unique action functional
$$S\doteq\int\limits_{M}d^4x\sqrt{|g|}L\doteq\int\limits_{M}d^4x\sqrt{|g|}\left[\frac{1}{2}
\psi^\dagger\left(D\psi\right)+\frac{1}{2}\left(D'\psi^\dagger\right)\psi\right].$$
A direct inspection of the above action shows us
that, up to a total derivative term, it is identical to the more common expression
$$S=\int\limits_{M}d^4x\sqrt{|g|}\psi^\dagger\left(-
\gamma^\mu\psi_{;\mu}+m\psi\right).$$
We define the (Hilbert) stress-energy tensor by the usual procedure, {\it i.e.},
$$T_{\mu\nu}\doteq\frac{2}{\sqrt{|g|}}\frac{\delta S}{\delta g^{\mu\nu}}.$$ 
An explicit realisation of this last identity in the case of spinor fields is much more involved due to the underlying orthogonal Lorentz frames whose explicit dependence on the metric must be accounted for. Nonetheless, a lengthy and, to a certain extent, tedious calculation, fully developed in \cite{Forger}, yields
\beq\label{SET}
T_{\mu\nu}=\frac{1}{2}\left(\psi^\dagger_{;(\mu} \gamma_{\nu)} \psi-\psi^\dagger\gamma_{(\mu}\psi_{;\nu)}\right)-Lg_{\mu\nu},
\eeq
where $()$ denotes idempotent symmetrisation; one should also notice that, being the field free, the Lagrangian vanishes on shell. 

If we contract \eqref{SET} with $g^{\mu\nu}$, we end up with the classical trace
$$
T\doteq g^{\mu\nu}T_{\mu\nu}=\frac{1}{2}\left(\psi^\dagger_{;\mu} \gamma^{\mu} \psi-\psi^\dagger\gamma_{\mu}\psi_;^{\mu}\right)
-4L=-m\psi^\dagger\psi,
$$
where, in the second equality, we have evaluated the left hand side on shell by means of \eqref{Dirac}. Hence, as
expected, the trace vanishes on shell for conformally invariant, {\it i.e.}, massless, Dirac fields.

If we consider instead the covariant conservation, we need to calculate 
\begin{gather}
\nabla^\mu T_{\mu\nu}
=\frac{1}{4}\left\{-\psi^\dagger_{;\nu}D\psi+\left[D'\psi^\dagger\right]_{;\nu}\psi-D'\psi^\dagger\psi_{;\nu}+\psi^\dagger\left[D\psi\right]_{;\nu}+P\psi^\dagger\gamma_\nu\psi-\psi^\dagger\gamma_\nu P\psi\right\}-L_{;\nu},\notag
\end{gather}
which vanishes on shell.

\subsection{The quantum stress-energy tensor: the problem} \label{quantum1} 

In the next step, we would like to define the quantum version of the stress-energy tensor for Dirac fields. Since we have a well-defined notion of Wick polynomials at hand, it would be easy to just take
the classical expression for the stress-energy tensor and replace the occurring field monomials with their normal ordered quantum counterparts. This way, one would easily get an element of $\cA_{ext}(M,g)$ which, GNS-represented with respect to a Hadamard state, would be a well-defined operator valued smooth function. As we will shortly see, however, this procedure would not yield a meaningful object. To understand this, let us take a slight detour and think about the properties we would like a quantum stress-energy tensor to have.

From the point of view quantum field theory over curved background, the most important entity to take into
account as a guide in the search for a good quantum stress-energy tensor is of course the semi-classical
Einstein's equation, {\it viz.},
\begin{gather}\label{semicla}
G_{\mu\nu}(x) = 8\pi G\omega(\wick{T_{\mu\nu}(x)}), 
\end{gather}
where $G_{\mu\nu}$ denotes the Einstein tensor $R_{\mu\nu}-\frac{1}{2}R g_{\mu\nu}$, $G$ is the gravitational
constant and $\wick{T_{\mu\nu}}$ is a suitable regularised expression for the quantum stress-energy tensor; this
equation thus describes the back-reaction of the quantum field on the background. The legitimate question
which now arises, is under which circumstances this equation makes sense at all. Regarding the {\it form} of
the equation, we will restrict ourselves to point out that it can be derived by formally expanding a quantum
metric and a quantum field about any classical vacuum solution of the Einstein's equation and computing the equation of motion for the expected metric while keeping only
``tree-level'' graviton contributions and ``loop-level'' quantum field contributions. Since one discards ``loop-level'' graviton contributions, the equation derived in this manner can only make sense as a model equation, or maybe for special states. We refer the interested reader to \cite{FW96} and the references cited therein for an exhaustive treatment of this topic while we shall continue dwelling upon the properties of the quantities appearing in (\ref{semicla}).

The first, and certainly obvious, observation is that we need a regularised expression of the stress-energy
tensor to obtain a finite expectation value, {\it i.e.}, a finite right hand side. The next observation is
that the left hand side of (\ref{semicla}) is a classical and ``sharp'' quantity, while the right hand side
is a probabilistic object. Such a situation can of course only be a sensible one if the fluctuations of the
probabilistic quantity involved are small in comparison to the quantity itself, finite in particular. These
considerations are exactly those which led to consider and select the states of Hadamard type as the physical
and reasonable ones among the myriad of states available in quantum field theory on curved spacetimes. As a
matter of fact, if we define normal ordering by means of the Hadamard singularity and we evaluate the
normal-ordered (and smeared) stress-energy tensor on Hadamard states, we automatically get a quantity with
finite fluctuations. This stems from the fact that powers of the Hadamard bidistribution are again
well-defined bidistributions, thanks to the special form of the Hadamard wavefront set. Regarding
quantitative statements about the fluctuations of the expected stress-energy tensor, it seems that in general
{\it a priori} statements are not possible and one has to look at solutions of the semiclassical Einstein's equations to {\it a posteriori} compute the fluctuations on these solutions and inspect to what extent these are to be trusted.

It will prove helpful to realise that $\omega(\wick{T_{\mu\nu}(x)})$ for a $\wick{T_{\mu\nu}(x)}$ given as some linear combination of the previously defined Wick monomials and evaluated on a Hadamard state $\omega$ can be equivalently expressed as
\beq\label{pointsplit}
\omega(\wick{T_{\mu\nu}(x)}) \doteq  \frac{1}{8\pi^2}Tr\left[D_{\mu\nu}(x,y)W(x,y)\right],
\eeq
where $D_{\mu\nu}$ is some (bi)differential operator specified by the choice of linear combination of Wick monomials in the definition of $\wick{T_{\mu\nu}(x)}$, $Tr$ denotes the trace over spinor indices and we refer the reader to appendix \ref{tools} for the explanation of the possibly unfamiliar notations that arise in the context of bispinorial entities and will be extensively used in the following. As we have already remarked, the obvious choice of expression for the stress-energy tensor in terms of Wick monomials will not turn out to the best one. In terms of the above defined differential operator, this means that the canonical version derived from the classical stress-energy tensor (\ref{SET}), $$D^{can}_{\mu\nu}\doteq -\widetilde D^{can}_{\mu\nu}D'_y\doteq-\frac{1}{2}\gamma_{(\mu}\left(\nabla_{\nu)}-g_{\nu)}^{\nu'}\nabla_{\nu'}\right)D'_y,$$ is not well suited for defining a sensible $\omega(\wick{T_{\mu\nu}(x)})$.

Since we have assured ourselves that a right hand side of (\ref{semicla}) obtained by expressing
$\omega(\wick{T_{\mu\nu}(x)})$ as (\ref{pointsplit}) is in principle well defined, we could seek for additional
physical and consistency requirements that lead to a potential refinement of that procedure, {\it i.e.}, to a
sensible choice of $D_{\mu\nu}$. Pursuing a comparable aim, Wald \cite{Wald2, Wald3} has set up five axioms
that a meaningful expected stress-energy tensor should fulfil. These proved to be a valuable
tool in {\it a posteriori} legitimating known stress-energy tensor regularisation schemes in curved
spacetimes and stating to which extent they may differ from one another without raising doubts about their
validity. For the convenience of the reader, we list them in the following.

\definizione{\label{axioms} We say that $\omega(\wick{T_{\mu\nu}(x)})$ fulfils (the strong version of) Wald's axioms if it has the following five properties:
\begin{enumerate}
\item Given two quasifree states $\omega_1$ and $\omega_2$, such that $\omega_1^{-}(x,y)-\omega_2^{-}(x,y)$ is a smooth bispinor, $\omega_1(\wick{T_{\mu\nu}(x)})-\omega_2(\wick{T_{\mu\nu}(x)})$ is equal to $Tr\left[\widetilde D^{can}_{\mu\nu}\left(\omega_1^{-}(x,y)-\omega_2^{-}(x,y)\right)\right]$.
\item $\omega(\wick{T_{\mu\nu}(x)})$ is locally covariant in the following sense: let $$\chi: (M_1,g_1,SM_1,\rho_1) \mapsto (M_2,g_2,SM_2,\rho_1),$$ $$\alp_\chi: \cF(M_1, g_1) \to \cF(M_2, g_2)$$ as in subsection \ref{funct}. If two states $\omega_1$ and $\omega_2$ on $\cF(M_1, g_1)$ and $\cF(M_2, g_2)$ are related by $\omega_1 = \omega_2 \circ \alp_\chi$, then $$\omega_2(\wick{T_{\mu_2\nu_2}(x_2)}) = \chi_*\left(\omega_1(\wick{T_{\mu_1\nu_1}(x_1)})\right),$$ where $\chi_*$ denotes the push-forward of $\chi$ in the sense of covariant tensors.
\item $\nabla^\mu\omega(\wick{T_{\mu\nu}(x)})$=0.
\item On Minkowski spacetime and in the Minkowski vacuum state $\omega_{Mink}$,   $\omega_{Mink}(\wick{T_{\mu\nu}(x)})$ = 0.
\item $\omega(\wick{T_{\mu\nu}(x)})$ does not contain derivatives of the metric of order higher than 2.
\end{enumerate}
$\omega(\wick{T_{\mu\nu}(x)})$ is said to fulfil the reduced version of Wald's axioms, if only the first four statements hold.
}

Wald has originally stated the axioms for scalar fields, while the version we give here is modified to be suitable for Dirac fields. We reckon that a few comments both on the origin and on the meaning of the single axioms might be helpful for a potential reader in order to understand their relevance: 

\vskip .2cm 

\noindent{\em 1.} In a given Fock-representation of the quantum field, the non-diagonal matrix elements of the formal unrenormalised stress-energy tensor operator in the ``mode basis'' are already finite, because their calculation only involves ``finite mode sums'', while the calculation of the diagonal matrix elements involves ``infinite mode sums'' \cite{Wald2, Wald3}. To regularise the formal stress-energy tensor operator, it is therefore only necessary to subtract an infinite part proportional to the identity operator, thus leaving the non-diagonal matrix elements unchanged. Axiom 1 amounts to require such a ``minimal'' regularisation. This axiom is also related to so-called {\it relative Cauchy evolution} of a locally covariant field \cite{BFV, Sanders};  since the functional derivative of the relative Cauchy evolution involves the commutator with the stress-energy tensor operator, one could reformulate this axiom on the operator level requiring that any regularisation prescription yields the same relative Cauchy evolution. If we consider Hadamard states, such that $\omega_i^{-}(x,y)$ is locally given by $-D'_y(H^-(x,y)+W_i(x,y))/(8\pi^2)$ for $i=1,2$, the requirement is equivalent to demanding that the differential operator used in (\ref{pointsplit}) is given by $D^{can}_{\mu\nu}$ plus a term which does not influence the state dependence of $\omega(\wick{T_{\mu\nu}(x)})$.

\vskip .2cm 

\noindent{\em 2.} Taking the locality principle of quantum field theory and the covariance principle of general relativity seriously, we would like to have a $\omega(\wick{T_{\mu\nu}(x)})$ which describes the back-reaction of the quantum field on the spacetime in a local and covariant way. In fact, this axiom seems to have been an inspiration towards the formulation of locally covariant QFT, as described in the seminal paper \cite{BFV}.

\vskip .2cm 

\noindent{\em 3.} This axiom basically points out a necessary condition for the well-posedeness of the semiclassical Einstein's equations; namely, since the geometric left hand side of (\ref{semicla}) is conserved due to the Bianchi identities, also the right hand side should vanish under the action of the covariant derivative.

\vskip .2cm 

\noindent{\em 4.} It is a sensible prerequisite of any regularisation scheme for a field theory on a curved background that it should be possible to read it as an ``extension'' of the standard normal ordering in Minkowski spacetime, but there are good reasons to skip this axiom, {\it cf.}, \cite{FullingBook} and the second remark after theorem \ref{tethe}.

\vskip .2cm 

\noindent{\em 5.} Wald originally proposed this axiom in a rather technical and more strict way \cite{Wald2, Wald3},
essentially requiring that  $\omega(\wick{T_{\mu\nu}(x)})$ does not depend on derivatives of the metric of order higher
than the first. The underlying motivation is rooted, on the one hand, in the request of well-posedness of the
Cauchy problem for the Einstein's equations even with a non vanishing source and, on the other hand, in the need for a sensible ``classical'' limit of the semi-classical Einstein's equations (see the enlightening discussion in Wald's original
paper \cite{Wald2, Wald3}). Wald himself had realised, however, that the strict version of this axiom could not even be satisfied in the classical theory and has, thus, proposed the weaker one stated here.  Unfortunately, further examinations have revealed that even this weaker version does not seem possible to fulfil in massless theories without introducing an artificial length scale into the theory; therefore, the axiom has been discarded. We still believe, however, that it could be fulfilled, though only under special circumstances. We shall comment on this issue at a later stage of the paper.

\vskip .2cm

Using these axioms, Wald could prove that a uniqueness result for $\omega(\wick{T_{\mu\nu}(x)})$ can be obtained.
The first two axioms already imply that the results from two different sensible regularisation schemes can
only  differ by a local curvature tensor. The third and fourth axiom then imply that this local curvature
tensor is conserved and vanishes if the spacetime is locally flat. Requiring that this term has the correct
dimension of $m^4$, the possible tensors are presumably only the ones obtained by varying a Lagrangian of the
form $$m^4\left(F_1\left(\frac{R}{m^2}\right)+F_2\left(\frac{R_{\mu\nu}R^{\mu\nu}}{m^4}\right)+F_3\left(\frac
{R_{\mu\nu\rho\tau}R^{\mu\nu\rho\tau}}{m^4}\right)\right)$$ with respect to the metric, with some
dimensionless functions $F_i(x)$. Requiring suitable analyticity properties with respect to the curvature tensors and $m$, in \cite{HW04}, it has been shown that the only possibilities are $m^4 g_{\mu\nu}$ ($F_1=1$)\footnote{A term proportional to the metric is not allowed if one seeks to fulfil the third axiom. As we will see later, however, it does not seem to be possible to fix this term in a way that is compatible with analyticity in $m$. Furthermore, the results of Hollands and Wald regarding the restriction of the possible regularisation freedom by demanding analytic dependence on curvature and mass have only been obtained for scalar fields. Since the stress-energy tensor for Dirac fields is an observable and thus still a ``scalar'' field, their results can be, nonetheless, presumably extended to this case.}, $m^2 G_{\mu\nu}$ ($F_1=x$) and the three local curvature tensors $I_{\mu\nu}$ ($F_1=x^2$), $J_{\mu\nu}$ ($F_2=x$), $K_{\mu\nu}$ ($F_3=x$), cf. appendix \ref{tools} for their full expression. In fact, we will later show that changing the scale $\lambda$ in the regularising
Hadamard bidistribution $H^-$ amounts to changing $\omega(\wick{T_{\mu\nu}(x)})$  exactly by a tensor of this
form and, furthermore, the attempt to regularise Einstein-Hilbert quantum gravity at one loop order
automatically yields a renormalisation freedom in form of such a tensor as well \cite{thooft}. Having in mind
how the semiclassical Einstein's equation may be derived, these two arguments are of course related by means of internal consistency.\footnote{In fact, at least in the case of scalar fields, the combination of the local curvature tensors appearing as the finite renormalisation freedom in \cite{thooft} is, up to term which seems to be an artefact the dimensional regularisation employed in that paper, the same that one gets via changing the scale in the regularising Hadamard bidistribution.} Using
the Gauss-Bonnet-Chern theorem in four dimensions, which states that  $$\int\limits_{M}d\mu(x)R_{\mu\nu\rho
\tau}R^{\mu\nu\rho\tau}-4R_{\mu\nu}R^{\mu\nu}+R^2$$ is a topological invariant and, therefore, has a vanishing
functional derivative with respect to the metric \cite{alty,thooft}, one can restrict the freedom even
further by removing $K_{\mu\nu}$ from the list of allowed local curvature tensors.

\subsection{The quantum stress-energy tensor: the solution and its trace anomaly} \label{quantum2} 

We now seek to exploit the above axioms in order to specify a sensible choice of differential operator $D_{\mu\nu}$.
Looking at our proposed regularisation procedure (\ref{pointsplit}), the first obstacle to overcome seems to be the covariant conservation axiom. As we have seen above, conservation of the stress-energy tensor in the classical case is a direct consequence of the equation of motion. Since we are regularising by subtracting from the two-point function the Hadamard bidistribution, which is in general not a solution of the Dirac equation(s), $D^{can}_{\mu\nu}$ applied to the thereby obtained smooth bispinor will in general not yield a conserved quantity. A viable solution to such problems calls for the modification of the classical stress-energy tensor (\ref{SET}) by terms which vanish on shell, while, at the same time, they help restoring covariant conservation on the quantum side. As in the case of scalar fields \cite{Mo03}, it seems that the only possible option is to add multiples of the Lagrangian to the classical expression of the stress-energy tensor. We thus propose the following classical stress-energy tensor as a starting point
$$
T_{\mu\nu}=\frac{1}{2}\left(\psi^\dagger_{;(\mu} \gamma_{\nu)} \psi-\psi^\dagger\gamma_{(\mu}\psi_{;\nu)}\right)+cLg_{\mu\nu}
$$ 
and we look for a $c\in\bR$ that yields a sensible $\omega(\wick{T_{\mu\nu}(x)})$. This is tantamount to the choice of the following differential operator for the point-splitting process (\ref{pointsplit})
\beq\label{dc}\begin{aligned} D^{c}_{\mu\nu}& \doteq D^{can}_{\mu\nu}-\frac{c}{2}g_{\mu\nu}\left(D'_x+D_y\right)D'_y\\&=-\frac{1}{2}\gamma_{(\mu}\left(\nabla_{\nu)}-g_{\nu)}^{\nu'}\nabla_{\nu'}\right)D'_y-\frac{c}{2}g_{\mu\nu}\left(D'_xD'_y-P_y\right).
\end{aligned}
\eeq

Before proceeding to prove that there indeed exists a suitable choice of $c$, we would like to anticipate another result, which can be easily understood from the aforementioned line of argument. Let us remember that there is another property of the classical stress-energy tensor stemming from the equations of motion: it has vanishing trace in the massless (and therefore conformally invariant) case. Following the above discussion, on the one hand, it seems that one might need to give up this property at a quantum level, while, on the other hand one, one could still hope that the choice of $c$ also provides a vanishing trace. Alas, it will turn out that this is not the case. One can only fix $c$ in a way such that $\omega(\wick{T_{\mu\nu}(x)})$ has vanishing trace for $m=0$, but conservation is inevitably spoilt. Since we have already realised that conservation is indeed an essential requirement for the right hand side of the semiclassical Einstein's equations, we will have to accept that $g^{\mu\nu}\omega(\wick{T_{\mu\nu}(x)})$ is not vanishing in the massless case. This goes under the name of {\it trace anomaly}.

\teorema\label{tethe}{Let $\lambda_m \doteq 2\exp(\frac{7}{2}-2\gamma)m^{-2}$ for $m\neq0$ and $\lambda_m$ arbitrary for $m=0$, where $\gamma$ denotes the Euler-Mascheroni constant, choose the Hadamard bidistribution to be the one with $\lambda=\lambda_m$ and let $\omega(\wick{T_{\mu\nu}(x)})$ be defined as in (\ref{pointsplit}), with the differential operator $D_{\mu\nu}=D^{-1/6}_{\mu\nu}$ defined as in (\ref{dc}). Then $\omega(\wick{T_{\mu\nu}(x)})$ fulfils the reduced version of Wald's axioms. Furthermore, it exhibits the following trace (anomaly)
\begin{align}\label{anomaly} g^{\mu\nu}\omega(\wick{T_{\mu\nu}(x)})=&-\frac{1}{\pi^2}\left(\frac{1}{1152}R^2+\frac{1}{480}\square R -\frac{1}{720}R_{\mu\nu}R^{\mu\nu}-\frac{7}{5760}R_{\mu\nu\rho\tau}R^{\mu\nu\rho\tau}\right)\\
&-\frac{1}{\pi^2}\left(\frac{m^4}{8} + \frac{m^2R}{48}\right) +mTr\left[D'_yW(x,y)\right].\notag\\
=&\frac{1}{2880\pi^2}\left(\frac72 C_{\mu\nu\rho\tau}C^{\mu\nu\rho\tau} +11\left(R_{\mu\nu}R^{\mu\nu}-\frac 13 R^2\right) -6\square R \right)\notag\\
&-\frac{1}{\pi^2}\left(\frac{m^4}{8} + \frac{m^2R}{48}\right) +mTr\left[D'_yW(x,y)\right].\notag\end{align}}

\proof{We begin by computing $\nabla^{\mu}\omega(\wick{T_{\mu\nu}(x)})$ and $g^{\mu\nu}\omega(\wick{T_{\mu\nu}(x)})$, leaving $c$ unspecified for the moment.
Applying Synge's rule and taking into account that $[C_{\mu\nu}, \gamma^{\mu}]=0$ and $[g_{\nu\,;\mu}^{\nu'}]=0$ ({\it cf.}, appendix \ref{tools}), we get
\beq\begin{aligned}
8\pi^2\nabla^{\mu}\omega(\wick{T_{\mu\nu}(x)}) & = \nabla^{\mu} Tr\left[D^c_{\mu\nu}(x,y)W(x,y)\right]= Tr\left[(\nabla^{\mu}+g^{\mu}_{\mu'}\nabla^{\mu'})D^c_{\mu\nu}(x,y)W(x,y)\right]\\
& = Tr\left[\left\{\frac{1}{4}\left(g_{\nu}^{\nu'}\nabla_{\nu'}-\nabla_{\nu}\right)\left(D'_xD'_y+P_y\right)+\frac{1}{4}\gamma_\nu D'_y(P_y-P_x)\right.\right.\\&\left.\left.\quad\qquad-\frac{c}{2}\left(g_{\nu}^{\nu'}\nabla_{\nu'}+\nabla_{\nu}\right)\left(D'_xD'_y-P_y\right)\right\}W(x,y)\right].
\end{aligned}\notag\eeq
Remembering that $-D'_y(H^-+W)$ is the local two-point
distribution of a state, it follows that $H^-+W$ is subject to the
distributional differential equations
$D'_xD'_y(H^-+W)=0=P_y(H^-+W)$. Thus, we can safely replace $W$
in the above equation by $-H^-$, since every appearing term involves one of
the two aforementioned differential operators. Such a procedure
yields \beq\begin{aligned} 8\pi^2\nabla^{\mu}\omega(\wick{T_{\mu\nu}(x)}) &
=
Tr\left[\left\{\frac{1}{4}\left(\nabla_{\nu}-g_{\nu}^{\nu'}\nabla_{\nu'}\right)\left(D'_xD'_y+P_y\right)+\frac{1}{4}\gamma_\nu
\left(\gamma^{\mu'}\nabla_{\mu'}+m\right)(P_x-P_y)\right.\right.\\&\left.\left.\quad\qquad-\frac{c}{2}\left(g_{\nu}^{\nu'}\nabla_{\nu'}+\nabla_{\nu}\right)\left(P_y-D'_xD'_y\right)\right\}H^-(x,y)\right].
\end{aligned}\notag\eeq
Now we can insert the various coincidence point limits of the differentiated Hadamard bidistribution $H^-$ computed in proposition \ref{cplimits} of appendix \ref{tools} to obtain
$$8\pi^2\nabla^{\mu}\omega(\wick{T_{\mu\nu}(x)})=-(1+6c)Tr[V_1(x,y)]_{;\nu}.$$
For the trace we use both the insights on the parallel transport of gamma matrices from appendix \ref{tools} and the arguments already employed in the computation of the conservation to get
\beq\begin{aligned}
8\pi^2g^{\mu\nu}\omega(\wick{T_{\mu\nu}(x)})&=g^{\mu\nu} Tr\left[D^c_{\mu\nu}(x,y)W(x,y)\right]\\
&=Tr\left[\left\{-\left(2c+\frac{1}{2}\right)\left(D'_xD'_y-P_y\right)+mD'_y\right\}W(x,y)\right]\\
&=Tr\left[\left(2c+\frac{1}{2}\right)\left(D'_xD'_y-P_y\right)H^-(x,y)+mD'_yW(x,y)\right]\\
&=-6(4c+1)Tr[V_1(x,y)]+mTr\left[D'_yW(x,y)\right].
\end{aligned}\notag\eeq
If we look at the above two results and we use the data from appendix \ref{tools}, we realise that, since $Tr[V_1(x,y)]$ is in general neither vanishing nor constant, we need to set $c=-1/6$ to assure conservation, thus yielding the asserted trace and, particularly, the trace anomaly in the massless case.\\
\indent Let us now proceed to check the validity of the first two axioms. From the above two calculations we can extract that the term we have added to the canonical differential operator $D^{can}_{\mu\nu}$ to achieve the looked-for differential operator $D^{c}_{\mu\nu}$ contributes to $\omega(\wick{T_{\mu\nu}(x)})$ a term proportional to $$g_{\mu\nu}Tr\left[\left(D'_xD'_y-P_y\right)H^-(x,y)\right] = -12g_{\mu\nu}Tr[V_1(x,y)].$$ This term is clearly independent of the state, {\it i.e.}, $W(x,y)$, thus, axiom 1 holds for our regularisation scheme and two Hadamard states $\omega_1, \omega_2$. Furthermore, since the Wick monomials are locally covariant quantum fields as discussed in the previous section, the same holds true for $\wick{T_{\mu\nu}(x)}$ and axiom 2 is straightforwardly fulfilled.\\
\indent What now remains to be shown is the vanishing of
$\omega(\wick{T_{\mu\nu}(x)})$ on Minkowski spacetime and in the
Minkowski vacuum state, provided we choose a suitable scale
$\lambda$ in the definition of $H^-$. In this setting, we have $\omega_{Mink}^-(x,y)=-D'_y
\omega_{2, Mink}^+(x,y)I_4$, with the scalar two-point function
$\omega_{2, Mink}^+$\footnote{Recall that, according to our definition, $\om^-$ is of ``positive frequency type''.} \cite{scharf}. The latter can be specified as
$$\omega_{2, Mink}^+(x,y)=\lim\limits_{\epsilon\rightarrow
0^+}\frac{4m}{(4\pi)^2\sqrt{2\sigma_{\epsilon}(x,y)}}K_1(m\sqrt{2\sigma_{\epsilon}(x,y)}),$$
where $K_1$ denotes a modified Bessel function, the expression is to
be understood in the sense of analytic continuation for negative
values of the geodesic distance, and the limit specifies how to
approach the branch cut of the squared root \cite{Mo03}. Expanding
this in terms of $\sigma$ (suppressing the $\epsilon$ for
simplicity) we have 
\begin{gather*}
8\pi^2\omega_{2,Mink}^-(x,y)=\left\{\frac{1}{\sigma}+\left(\frac{m^2}{2}+\frac{m^4}{8}\sigma+f_1(\sigma)\sigma^2\right)\ln\left(\frac{e^{2\gamma}m^2\sigma}{2}\right)+\right.\\
\left.-\frac{m^2}{2}\left(1+\frac{5m^2\sigma}{8}\right)+\sigma^2f_2(\sigma^2)\right\}I_4,
\end{gather*}
where the $f_i$ appearing in this paragraph are smooth functions. In
Minkowski spacetime, the Dirac Hadamard bidistributions $H^{\pm}$ are
simply the scalar ones times the unit matrix. This is not
surprising since, as visible in appendix \ref{tools}, the nontrivial
matrix part of the Dirac Hadamard coefficients stems from the
curvature of the spin connection, which vanishes in flat spacetimes. We thus have
$$8\pi^2H_{Mink}^-(x,y)=\left\{\frac{1}{\sigma}+\left(\frac{m^2}{2}+\frac{m^4}{8}\sigma+f_1(\sigma)\sigma^2\right)\ln\left(\frac{\sigma}{\lambda^2}\right)\right\}I_4.$$ 
If we take into account this singular part, a short computation yields
$$8\pi^2W_{Mink}(x,y)=\left\{\frac{m^2}{2}\left(\ln\left(\frac{e^{2\gamma}m^2\lambda^2}{2}\right)-1\right)+\left(\frac{m^4}{8}\ln\left(\frac{e^{2\gamma}m^2\lambda^2}{2}\right)-\frac{5m^4}{16}\right)\sigma+f_3(\sigma)\sigma^2\right\}I_4$$ and we can straightforwardly compute
$$\omega_{Mink}(\wick{T_{\mu\nu}(x)})=-\frac{m^4}{8}g_{\mu\nu}\left(\ln\left(\frac{e^{2\gamma}m^2\lambda^2}{2}\right)-\frac{7}{2}\right),$$
which vanishes for $\lambda = \lambda_m =
2\exp(\frac{7}{2}-2\gamma)m^{-2}$. In the massless case, both
$H^-_{Mink}$ and $\omega_{2, Mink}$ only consist of the
$\sigma^{-1}$ term, such that $\omega_{Mink}(\wick{T_{\mu\nu}(x)})$ is
trivially vanishing, independent of any scale $\lambda$. }

It is interesting to notice that the above analysis seems to suggest that the common idea which associates the emergence of the trace anomaly to the conservation of the stress-energy tensor is somehow inappropriate since such anomaly seems rooted in the loss of the equations of motion at a quantum level alone. As a matter of fact the needed modification of the classical stress-energy tensor does not {\it evoke} the trace anomaly, but it only {\it modifies} it. 

To conclude the section, a few further remarks are in due course:
\remark{Our result for the trace anomaly coincides with previous ones obtained by means of gravitational index theorems \cite{chris1} and point-splitting techniques \cite{Christensen}. Both approaches have made use of the {\it DeWitt-Schwinger expansion}, which can not be defined rigorously \cite{Wald79, Brown, FullingBook, Mo03}. Moreover, even if this expansion reproduces the Hadamard singularity structure, it seems that calculations are much shorter if one expresses it directly through the Hadamard series. The previous attempts had, however, one advantage, namely, they had expressed the expected stress-tensor as the functional derivative of a (diffeomorphism-invariant) effective action, such that the result has been manifestly conserved. The nice discussion which follows definition 2.1 of \cite{Mo03} gives an explanation of how the extra term in our derivative operator $D^c_{\mu\nu}$ can be understood in this context.}

\remark{By changing the scale $\lambda$ in the Hadamard bidistribution $H^-$ to $\lambda'$, one modifies the (definition of the) smooth part $W$ by $2V\ln{\lambda'/\lambda}$, such that $\omega(\wick{T_{\mu\nu}(x)})$ changes by a term proportional to $Tr[D^c_{\mu\nu}V]$. Since we know from proposition \ref{extrasmooth} that $D'_yV$ fulfils both Dirac equations of motion, we can {\it a priori} deduce that this term is automatically conserved and furthermore traceless in the conformal case. Thus, it follows that both the determination of the correct $c$ to be inserted in $D^c_{\mu\nu}$ and the trace anomaly are independent of the scale $\lambda$. Even if we already know the properties of $Tr[D^c_{\mu\nu}V]$ beforehand, it is enlightening to calculate its explicit form. The result is $$Tr[D^c_{\mu\nu}V]=\frac{m^4}{2}g_{\mu\nu}-\frac{m^2}{6}G_{\mu\nu}+\frac{1}{60}\left(I_{\mu\nu}-3J_{\mu\nu}\right),$$
where the linear combination of $I_{\mu\nu}$ and $J_{\mu\nu}$ appearing in the above formula is traceless, {\it cf.}, appendix \ref{tools}. This term is well within and even exhausts the regularisation freedom discussed after definition \ref{axioms}.\\\indent
Furthermore, one should recall that, in the massive case, $\lambda$ had to be fixed in terms of inverse powers of $m$ to assure vanishing of $\omega(\wick{T_{\mu\nu}(x)})$ in Minkowski spacetime. Hence, if one demands continuous dependence of $\omega(\wick{T_{\mu\nu}(x)})$ on $m$, it does not seem possible to fulfil the third axiom of definition \ref{axioms} in this way.}

\remark{ Even if we are able to fix the scale $\lambda$, we could in principle still
add multiples of $m^2G_{\mu\nu}$, $I_{\mu\nu}$, and $J_{\mu\nu}$ to $\omega(\wick{T_{\mu\nu}(x)})$ without spoiling
the validity of the first four of Wald's axioms, though possibly modifying the $\square R$ term in the trace anomaly. This freedom can be derived in a more general sense, by already viewing all Wick products as being uniquely defined only
up to terms depending on the mass and the local curvature, where the possible regularisation freedom is
partially restricted by suitable consistency conditions, {\it e.g.} ``Leibnitz rules''. This approach has been
developed and pursued successfully by Hollands and Wald in \cite{HW04} and has the advantage to also
encompass interacting fields. A treatment along their lines will, as already remarked in subsection \ref{quantum1}, presumably yield the same renormalisation freedom for the stress-energy tensor like the one found here, as it happens in the scalar case \cite{HW04}.\\\indent
Furthermore, the arising of the conserved local tensors $I_{\mu\nu}$ and $J_{\mu\nu}$ puts us in the position to understand why the fifth of Wald's axioms in definition \ref{axioms} is problematic. Let us consider the massive case, where we can fix the scale $\lambda$. Since $I_{\mu\nu}$ and $J_{\mu\nu}$ contain terms involving fourth order derivatives of the metric, we may hope to cancel terms of that type occurring in $\omega(\wick{T_{\mu\nu}(x)})$ by adding a fixed linear combination of those two tensors. In the massless case, however, there is, up to our knowledge, no physically sensible way to fix $\lambda$. Therefore, one has no control on the multiples of $I_{\mu\nu}$ and $J_{\mu\nu}$ occurring in $\omega(\wick{T_{\mu\nu}(x)})$ and thus no way to cancel them.\\\indent
Nonetheless, there are scenarios where the situation with respect to the mentioned axiom is not that
pernicious. On cosmological, {\it i.e.}, Friedmann-Robertson-Walker backgrounds, the semi-classical
Einstein's equations (\ref{semicla})  can be reduced to an equation for the traces of both sides plus a
conservation equation for the right hand side \cite{dfp}. Thus, it seems that one has the chance to fulfil
the fifth axiom for both massive and massless fields in this simplified setting, since a change of scale does
not add fourth order derivative terms to the trace of the expected stress-energy tensor. In fact, as already
explained in the introduction, this observation has been used in \cite{dfp} to obtain stable solutions of the
semi-classical Einstein's equation at late times.}

\se{Conclusions and outlook}

We have extensively discussed the structure of free Dirac fields both at a classical and at a quantum level.
While, in the first case, we have mostly reviewed standard approaches, in the latter scenario we have
achieved a twofold goal. Particularly, we have started the discussion of quantised Dirac spinors by
exploiting the selfdual framework introduced by Araki which treats spinor and cospinor fields as a combined
single object and allows to formulate the quantisation procedure in a locally covariant way. This step has
been fully undertaken by Sanders for the first time and we have recalled the essential steps and features of
this construction. Employing already available and known properties of Hadamard states, we have subsequently
been able to introduce the extended algebra of Wick polynomials, the topic of section \ref{wicks} and the
first of our main results. As a second one, we have shown that, as in the scalar case, a physically sensible definition of the stress-energy tensor for Dirac fields on a curved background in terms of Wick polynomials is indeed possible with just one caveat: one has to add to the classical expression a suitable term which vanishes on-shell and hence does not alter classical  dynamics to obtain a conserved stress-energy tensor on the quantum side. Some new insights on Diracian Hadamard forms have constituted a prerequisite of this result, while one of its consequences is the emergence of a non-vanishing quantum trace of the stress-energy tensor, even if its value at a classical level is zero in the conformally invariant case. This result, which goes under the name of trace anomaly, has been previously known, but only as a result of formal calculations; it is thus derived here rigorously for the first time. 

On the overall, we reckon that this paper accomplishes also a further task, namely, it adds the insight that it is possible, interesting, but by no means straightforward to recast many of the already known rigorous results for scalar fields also for the spinor ones. Furthermore, our analysis opens several interesting questions to be tackled in future lines of research: the first one, which arises out of  section \ref{funct}, concerns the possibility to prove the time slice axiom for the extended algebra of fields (as well as for interacting field theories) in the scenario considered in the paper. If one follows the path paved in the scalar case in \cite{Chilian}, a positive answer seems definitively within our grasp. A further interesting problem originates from section \ref{SHS} in which Hadamard states are introduced and discussed; the Hadamard coefficients appearing in the singularity structure of such states are smooth bispinors and the question arises if their most remarkable feature in the scalar case, namely, their symmetry as proved in \cite{Moretti2, Moretti}, also appears in the spinor scenario. Such a property would be desirable since, for example, it would lead to many simplifications in the demanding calculations necessary in the construction of the conserved stress-energy tensor.  Although there are hints pointing towards this direction (see also \cite{SahlmannVerch}), we are far from a complete proof of such a symmetry and we thus feel this would be another rather interesting problem to tackle in the very next future.

Besides these rather formal lines of research, our results have also some remarkable consequences at a physical level. On the one hand we are now ready to answer the question posed in the introduction on the robustness of the results in \cite{dfp}; preliminary considerations seem to point towards this direction, though we leave a definitive answer to a future analysis. On the other hand, since our approach allows us to control the behaviour of free Dirac fields at a cosmological level, it is interesting to point out that free or perturbatively self-interacting fields with half-integer spin in cosmology arise in many models, such as baryogenesis through leptogenesis, where they often play a pivotal role. In these scenarios there are still many open questions to be answered and it seems that, often, the role of spacetime curvature effects are a priori discarded as negligible. Our experience suggests that this approximation might be too crude and, therefore, we would like to investigate these models in more detail in the framework of quantum field theory in curved spacetimes with the hope that such an analysis might lead to new and interesting physical consequences. 

\section*{Acknowledgements.} 

The work of C.D. is supported by the von Humboldt Foundation, that of T.H. by the German DFG Graduate School GRK 602, whereas N.P. gratefully acknowledges support by the German DFG Research Program SFB 676. We would like to thank K. Fredenhagen, V. Moretti, and R. Punzi for useful discussions. T.H. is especially grateful to R. Punzi for suggesting \cite{Ricci} to him. 


\appendix

\section{Useful tools and necessary calculations}\label{tools}

The aim of the appendix is to recollect, to clarify, and, occasionally, to also prove useful formulas which are needed in the main body of the paper and which are subject to potential ambiguities. These are often perniciously leading to potentially grievous sign mistakes or misunderstandings of a sort which we wish to hold off from a potential reader. 

\subsection{Notations, conventions, identities}\label{NC}

As a starting point, we would like to recollect our basic conventions regarding some symbols, whose exact definition often varies among the literature. In accord with section \ref{classical}, we work with spacetimes thought as four-dimensional, Hausdorff, smooth manifolds\footnote{As proven in \cite{Geroch}, second countability is automatically fulfilled for a four-dimensional, Hausdorff, smooth manifold of Lorentzian signature.}, endowed with a Lorentzian metric $g_{\mu\nu}$ with signature $(-,+,+,+)$. At the same time, other notable geometric quantities, namely, the Riemann and the Ricci tensor as well as the Ricci scalar,
are defined  via their components as follows
$$v_{\alpha;\be\ga} - v_{\alpha;\ga\be}\doteq
R_{\alpha\,\,\beta\gamma}^{\,\,\,\lambda} v_\lambda,\qquad
R_{\alpha\beta}\doteq
R_{\alpha\,\,\beta\lambda}^{\,\,\,\lambda},\qquad R\doteq
R_{\,\,\alpha}^\alpha,$$ 
where $v_\alpha$ are the components of an arbitrary covector; the extension to vectors and tensors of higher rank is then straightforward. As a last remark, we underline that the Riemann tensor possesses the symmetries
$$R_{\alpha\beta\gamma\delta}=-R_{\beta\alpha\gamma\delta}=-R_{\alpha\beta\delta\gamma}=R_{\gamma\delta\alpha
\beta}$$ and fulfils $$R_{\alpha\beta\gamma\delta}+R_{\alpha\delta\beta\gamma}+R_{\alpha\gamma\delta\beta}=0.$$ Finally, we define the Weyl tensor as it is usually done by the following expression $$C_{\alp\be\ga\de}=R_{\alpha \beta \gamma \delta}-\frac{1}{6}\left(g_{\alpha \delta}g_{\beta \gamma}+g_{\alpha \gamma}g_{\beta \delta}\right)R-\frac{1}{2}\left(g_{\beta \delta}R_{\alpha \gamma}+g_{\beta \gamma}R_{\alpha \delta}+g_{\alpha \delta}R_{\beta \gamma}+g_{\alpha \gamma}R_{\beta\delta}\right).$$

Similarly, we need to cope with geometric quantities related to the spin structure, introduced in definition \ref{spinstr}. Most of these are constructed out of the so-called $\ga$-matrices which satisfy the standard anticommutation relations \eqref{gammamu}, {\it i.e.}, $\left\{\gamma_\mu,\gamma_\nu\right\}=2g_{\mu\nu}$. Our choice of the metric signature entails that the $\ga$-matrices are different from the standard ones employed in quantum field theory books by an overall multiplicative factor $\pm i$. Consistently also with \eqref{gammas}, we stick to $+i$ and, therefore, the Dirac operator appearing in the Dirac equation for spinors becomes $D\doteq-\displaystyle{\not}\nabla+m$, whereas the operator nullifying a dynamically allowed cospinor is $D'\doteq\displaystyle{\not}\nabla+m$.

That said, apologising in advance for assigning the letter $C$ to two different objects, we define the components of the curvature tensor $C$ of the spin connection as $$V_{A;\be\ga} - V_{A;\ga\be }\doteq C_{A\,\,\beta\gamma}^{\,\,\,B} V_B, $$ where $V_A$ are the components of an arbitrary cospinor; as previously, the extension of this definition to spinors, also of higher rank, as well as of that for the Riemann and the spin curvature tensor in presence of mixed spinor-tensors, is straightforward. It follows from lemma \ref{covdev} that the
relation between the two curvature tensors is
$$C_{\,\,\,B\mu\nu}^{A}=\frac{1}{4}R_{\mu\nu\alpha\beta}\gamma^{\alpha A}_{\,\,\,\,\,\,C}\gamma^{\beta C}_{\,
\,\,\,\,\,B}.$$ Thus, $C$ possesses the
symmetries $$C_{AB\mu\nu}=-C_{BA\mu\nu}=-C_{AB\nu\mu}.$$ 

We also use the notational convention that a matrix acts from the left on spinors and from the right on cospinors,
{\it e.g.}, we resolve the Dirac operators as
$$D'\psi^\dagger=\psi^\dagger_{;\mu}\gamma^\mu+m\psi^\dagger,\qquad D\psi= -\gamma^{\mu}\psi_{;\mu}+m\psi.$$
If one strictly sticks to such convention, spinor indices can be safely suppressed, as we have already done in
the main body of the paper and as we will often do in the remainder of this appendix.

To conclude this subsection, we point out a few useful identities between the objects we have previously introduced. Starting from the gamma matrices, the product of an odd number of them has a vanishing trace. At the same time, if we consider an even number
\beq
Tr\,\ga_\mu\ga_\nu = 4 g_{\mu\nu},\quad Tr\,\ga_\mu\ga_\nu\ga_\alp\ga_\be = 4(g_{\mu\nu}g_{\alp\be}-g_{\mu\alp}g_{\nu\be}+g_{\mu\be}g_{\nu\alp})\notag\\
Tr\,\ga_{[\alp}\ga_{\be]}\ga_{[\ga}\ga_{\de]}\ga_\ep\ga_\ph =
4(g_{[\alp\ph}g_{\be][\ga}g_{\de]\ep}+g_{[\alp\ep}g_{\be][\de}g_{\ga]\ph}+g_{[\alp\de}g_{\be]\ga}g_{\ep\ph})\notag,
\eeq where $[\quad]$ here denotes idempotent antisymmetrisation. Furthermore, 

$$\ga^\mu\ga_\mu=4I_4, \quad \ga^\mu\ga^\alp\ga_\mu=-2\ga^\alp, \quad \ga^\mu\ga^\alp\ga^\be\ga_\mu=4g^{\alp\be}I_4,$$
\beq\label{6gamma}
\quad \ga^\mu\ga^\alp\ga^\be\ga^\ga\ga_\mu=-2\ga^\ga\ga^\be\ga^\alp, \quad \ga^\mu\ga^\alp\ga^\be\ga^\ga
\ga^\de\ga_\mu = 2(\ga^\de\ga^\alp\ga^\be\ga^\ga+\ga^\ga\ga^\be\ga^\alp\ga^\de).
\eeq
The last equalities we shall need are 
$$\ga^\alp C_{\alp\be} = C_{\alp\be}\ga^\be = \frac{1}{2}R_{\alp\be}\ga^\be, \quad [C_{\alp\be}, \ga_\ga]=R_{
\alp\be\rho\ga}\ga^\rho, \quad Tr\, C_{\alp\be}\ga_\ga\ga_\de = -2 R_{\alp\be\ga\de},\quad C_{\alp\be;}^{\,\,
\,\,\,\,\alp\be}=0,$$
$$Tr\,C_{\alp\be}C^{\alp\be}=-\frac{1}{2}R_{\alp\be\ga\de}R^{\alp\be\ga\de},\quad Tr\,C_{\alp\be}C^{\alp\be}\ga_\rho\ga_\tau=Tr\,C_{\alp\be}C^{\alp\be}g_{\rho\tau},$$ where the equalities not involving a trace can be proved by combining the symmetry properties of the Riemann tensor with the anticommutation relations of the $\ga$-matrices, {\it e.g.}, \begin{align}C_{\alp\be}\ga^\be&=\frac14 R_{\alp\be\rho\tau}\ga^\rho\ga^\tau\ga^\be=\frac14 (R_{\alp\rho\be\tau}+R_{\alp\tau\rho\be})\ga^\rho\ga^\tau\ga^\be=\frac14 R_{\alp\be\rho\tau}(\ga^\be\ga^\tau\ga^\rho+\ga^\rho\ga^\be\ga^\tau)\notag\\&=\frac14 R_{\alp\be\rho\tau}\left(-\ga^\rho\ga^\tau\ga^\be+2g^{\be\tau}\ga^\rho-2g^{\be\rho}\ga^\tau-\ga^\rho\ga^\tau\ga^\be+2g^{\be\tau}\ga^\rho\right)\notag\\&=\frac32 R_{\alp\rho}\ga^\rho - 2C_{\alp\be}\ga^\be\notag\\
\Leftrightarrow C_{\alp\be}\ga^\be &= \frac12 R_{\alp\be}\ga^\be.\notag\end{align}

\subsection{On the calculus of bispinor-tensors}

The notion of bispinor-tensors heuristically boils down to consider objects which contemporary transform as spinor-tensors at two spacetime points. In a more sound language, they are sections of an outer tensor product $VM \boxtimes WN$ of two vector bundles $VM$, $WN$ respectively over $M$ and $N$. $VM \boxtimes WN$ is nothing but a vector bundle over $M\times N$ with, calling $V$ and $W$ the typical fibres of $VM$ and $WN$, $V\otimes W$ as a typical fibre. Such a construction may seem awkward, but, in case $M=N$, it is indeed more fundamental than the familiar tensor product bundle $VM \otimes WM$, the latter being constructed out of $VM \boxtimes WM$ by pulling back via the map $M\ni x \mapsto (x,x)\in M\times M$.

For simplicity we will choose to collect all possible (bi)spinor-tensorial objects under the name of (bi)tensor, except in special case where we want to stress the character of the involved vector spaces. The bitensors occurring in this work are all defined only on a convex normal neighbourhood, since we need a unique
geodesic to connect the two points our bitensors depend on. We use
unprimed indices to indicate components stemming from tensorial
properties at $x$ and primed indices for those rooted in such
properties at $y$. Furthermore, we shall use the bracket notation
introduced by Synge to denote coincidence point limits of
bitensors, namely, $$[B(x,y)] \doteq \lim\limits_{y\to x} B(x,y),$$
where $B$ is some smooth bitensor, such that the limit is well
defined.

Let us now recall the bitensors used in this work and examine their properties. We will only mention the basic points while we refer to the works of DeWitt and Brehme, Fulling, Christensen, and to the review by Poisson for further, more exhaustive, details \cite{DeWitt, FullingBook, Christensen0, Poisson}.

As a starting point, we consider the halved squared geodesic distance $\sigma(x,y)$ taken with sign, sometimes also called {\it
Synge's world function}. Even if the geodesic distance itself might not be globally
smooth, it is such on geodesically convex normal neighbourhoods (provided smoothness of the metric) and it furthermore fulfils $\sigma_{;\mu}\sigma^{;\mu} = 2\sigma$, an identity which can be either explicitly computed or derived from
geometric considerations. In the following, we will, as it is
customary, drop the semicolon when indicating covariant derivatives
of $\sigma$. The aforementioned equation together with
$[\sigma]=[\sigma_\mu]=0$ and $[\sigma_{\mu\nu}] = g_{\mu\nu}$,
two identities arising out of the defining properties of the geodesic distance, completely suffice to determine $\sigma$, as well as all the properties we need, namely, the coinciding point limits of its higher derivatives. These can be obtained by means of an inductive procedure; as an example, in the case of $[\sigma_{\mu\nu\rho}]$, one differentiates
$\sigma_\mu\sigma^\mu = 2\sigma$ three times and then takes the coinciding point limit. Together with the already known relations, one obtains $[\sigma_{\mu\nu\rho}]=0$. At a level of fourth derivative, a new feature enters the fray, namely, one gets a linear combination of three coinciding fourth derivatives, though with different index orders. To relate those, one has to commute derivatives to rearrange the indices in the looked-for fashion, and this ultimately leads to the appearance of Riemann curvature tensors, {\it i.e.}, 
$$[\sigma]=[\sigma_\mu]=[\sigma_{\mu\nu\rho}]=0,\quad[\sigma_{\mu\nu}]=g_{\mu\nu},
\quad
[\sigma_{\mu\nu\varrho\tau}]=-\frac{1}{3}(R_{\mu\varrho\nu\tau}+R_{\mu\tau\nu\varrho}).$$
We stress that the discussion of these few identities is indeed much more valuable than just yielding the stated results since a potential reader is now able to calculate coinciding point limits both of derivatives of arbitrarily high order and of any bitensor; this holds true provided he is given the limits of lower order derivatives, an equation relating them to the higher ones, as well as the information of appropriate curvature tensors. We would like to remark at this point that, since one is ultimately interested in the coinciding point limits of certain bitensors most of the time, the in between computational steps often only require the knowledge of coinciding point limits of hierarchically lower objects, in contrast to having the necessity to know their full form. 

The next interesting bitensor is that of parallel transport
along a geodesic, an object depending both on the underlying vector bundle and on the considered linear connection. We will denote the parallel transport relating the tangent spaces at $x$ and $y$ as $g^\mu_{\nu'}$,
while the one relating spinors at those same points is denoted as $I^A_{B'}$.
With them at hand, parallel transporting a spinor-tensor $T=T^{A\mu}\,E_A\otimes \partial_\mu$ along
the geodesic connecting $y$ to $x$ amounts to the following identity
$$T^{A\mu} =I^A_{B'}g^\mu_{\nu'}T^{B'\nu'},$$ 
and a similar rule applies to higher spinor-tensors. One can reverse the role of $x$ and $y$, introducing the
inverses of the above two parallel transports, say $g^{\mu'}_{\nu}$ and ${I^{-1}}^{A'}_{B}$. On a practical ground, the construction of these two quantities boils down to finding a solution of the following partial differential equations:
$$g^\mu_{\nu';\alpha}\sigma^\alpha=I^A_{B';\alpha}\sigma^\alpha=0\quad\mbox{and}\quad[g^\mu_{\nu'}]=g^\mu_\nu,\quad[I^A_{B'}]={I_4} ^A_{B},$$ 
being $I_4$ the $4\times 4$ identity matrix. These identities, together with the properties of  $\sigma$
and the inductive procedure described at the beginning of this
paragraph, allow us to explicitly compute the derivatives of the parallel transports, the lowest ones being
\beq\label{paracpl}
[g^\mu_{\nu';\alpha}]=[I^A_{B';\alpha}]=0,\quad[g^\mu_{\nu';\alpha\beta}]=\frac{1}{2}R^\mu_{\,\,\,\nu\alpha\beta},\quad[I^A_{B';\alpha\beta}]=\frac{1}{2}C^A_{\,\,\,B\alpha\beta}.
\eeq
We shall henceforth suppress spinor indices, taking care to follow the afore described conventions, and, to conclude the section, we would like to point out the special parallel transport properties of both $\sigma$ and the gamma matrices. For the former we have, due to its geometric meaning,
$$g^{\mu'}_{\nu}\sigma_{\mu'} = -\sigma_{\nu},$$ 
whereas, for the latter, being covariantly constant, we have
$$I\gamma_{\mu'}I^{-1}g^{\mu'}_{\nu} = \gamma_{\nu}.$$

In this paper, we need to cope with the coinciding point limits of bitensors differentiated at both $x$ and $y$. The first, and maybe obvious, related statement is that derivatives at different points commute, so that we can always rearrange derivative indices in such a way that the unprimed ones are always on the left whereas the primed ones are always on the right. As a subsequent step, one notices that mixed coinciding point limits can be also calculated out of inductive paths. If one has the knowledge of the coinciding derivatives at the point $x$, however, one can extend it to those at $y$ by means of {\it Synge's rule}:
\lemma{Let $T$ be a smooth bitensor of arbitrary order; then its covariant derivatives possess the following property in the coinciding point limit (here suppressing all unessential indices):
$$[T_{;\mu'}]=[T]_{;\mu} - [T_{;\mu}].$$}

This has been proven by Synge for $\sigma$ exclusively, while, for the proof of an extension to arbitrary bitensors, one can refer to section 2.2 in Poisson \cite{Poisson} or to Christensen \cite{Christensen0}.

\subsection{On the Hadamard recursion relations and related results}\label{Hadcoeff}

As we have seen in section \ref{SHS}, in order to ``construct'' the two-point functions $\om^\pm(x,y)$ of a Hadamard state, we need to specify the distribution kernels $H^\pm(x,y)$ and the smooth bispinor $W(x,y)$, which must satisfy

$$P_xH^\pm(x,y)\in\cE(DM\otimes D^*M), \quad P_yH^\pm(x,y)\in\cE(DM\otimes D^*M)$$ and $$D'_xD'_y\left(H^\pm(x,y)+W(x,y)\right)=P_y\left(H^\pm(x,y)+W(x,y)\right)=0.$$ From this it follows that $D'_xD'_y H^\pm(x,y)$ is a smooth bispinor as well. Furthermore, due to proposition \ref{extrasmooth}, there are even more differential operators, which, applied to $H^\pm(x,y)$, yield a smooth bispinor. Let us collect them all in the following: 
\beq\label{Px}
P_x =-D'_xD_x,\quad P_y = -D'_yD_y,\quad D'_xD'_y, \quad D_xD_y \quad
\mbox{and}\quad D_x-D'_y=D_y-D'_x.
\eeq The aim of this section is to use these data to determine the Hadamard bidistributions $H^\pm(x,y)$ and to calculate the various coinciding point limits of their derivatives which are necessary for the proof of theorem \ref{tethe}. Following the path paved in the preceding sections, let us recall that the index structure of $\om^\pm$ is $$\om^\pm(x,y)=\om^\pm(x,y)_A^{\;\;\;B'}\, E^A(x)\otimes E_{B'}(y),$$ 
and that $H^\pm$ and $W$ inherit this structure, and let us suppress spinor indices in the following.

 Although $H^\pm(x,y)$ and $W(x,y)$ are bispinors, we recall from the main body of the paper that their form slavishly mimics that of the kernels specifying the two-point function in the theory of scalar fields, {\it viz.}, 
\begin{gather}
H^\pm(x,y)=\frac{U(x,y)}{\sigma_{\pm\epsilon}(x,y)}+V(x,y)\ln\frac{\sigma_{
\pm\epsilon(x,y)}}{\lambda^2},\label{utile}\\
V(x,y)\doteq\sum\limits_{n=0}^\infty V_n(x,y)\sigma(x,y)^{n},\label{Vexp2}\\ W(x,y)\doteq\sum\limits_{n=0}^\infty W_n(x,y)\sigma(x,y)^{n},\label{Wexp2}
\end{gather}
where $\sigma_{\pm\epsilon}(x,y)\doteq \sigma(x,y)\pm 2i\epsilon\left(T(x)-T(y)\right)+\epsilon^2,$ with $\epsilon>0$ and $T$ being a temporal function whose existence is guaranteed since the background is globally hyperbolic
\cite{Bernal, Bernal2}. As already commented in the main text, $\lambda$ is a reference distance employed to make the argument of the logarithm dimensionless, while the remaining objects, the so-called {\it Hadamard coefficients} $U$ and $V$, are smooth bispinors. As we will see shortly, $U$ as well as $V$ depend only on the geometry of the underlying background and the mass, whereas $W$ fully characterises the state, namely, the two-point functions of two Hadamard states differ only by a smooth function and such a difference is indeed encoded in $W$. 

To determine $U$, $V$, and $W$, we need to use the knowledge on the differential operators \eqref{Px} which, once applied to $H^\pm$, give smooth bispinors. To make the following formulas more readable, we choose to omit the regularising $\ep$ - and thus the $\pm$ index of $H^\pm$ -, the reference length $\la$, and the dependence of the kernels on the spacetime points $(x,y)$. That said, we can in principle take either of the second order differential operators listed in \eqref{Px} to recursively calculate $U$ and $V$; we will employ $P_x$, as this is familiar from the computations in the scalar case. 

Applying $P_x$ to $H$, we obtain potentially singular terms proportional to $\sigma^{-n}$ for $n=1,2,3$ and to $\ln \si$ as well as smooth terms proportional to positive powers of $\si$. We know, however, that the total result is smooth and one possible way to achieve this is to demand that the coefficients of the potentially singular terms are identically vanishing. Let us stress that, since we do {\it a priori} not know if $U$ contains positive powers of $\si$, the terms proportional to negative powers of $\si$ could in principle cancel each other to yield a smooth result. It is therefore a choice and not a necessity to require the coefficients of the inverse powers of $\si$ to vanish, and it is, furthermore, not guaranteed that the result of this procedure does not depend on the choice of the second order differential operator out of the possible ones listed in \eqref{Px}. The afore laid down line of argument does, however, not hold for the coefficients of $P_x H$ proportional to $\ln \si$; since $U$ and $V$ are required to be smooth, they can not contain a logarithmic dependence on $\si$ and the terms proportional to $\ln \si$ have to vanish necessarily.

The result of the previously described procedure are the so-called {\it Hadamard recursive relations}, which, in the scalar case, have been studied by several authors (see for example \cite{Moretti}). In the case of Dirac fields, there are results on the coinciding point limits of the Hadamard coefficients up to $V_1$ computed in \cite{Christensen}; the form of the Hadamard singularity employed in this work is, however, a different one  related to the non-rigorous DeWitt-Schwinger expansion, but formally, the relation between the different recursion relations arising in the two constructions is well-known.

After having discussed the Hadamard recursion relations, we shall show  how they arise explicitly. Let us thus examine the terms $U/\si$ and $V\ln \si$ individually. Starting with the latter, we have
\begin{gather*}
P_x( V\ln\si)=(P_xV)\ln\sigma+
\sum\limits_{n=0}^\infty\left(V_n(\square_x\sigma-2+4n)+2\sigma^\mu V_{n;\mu}\right)\sigma^{n-1},
\end{gather*}
where we have employed the identity $\sigma^\mu\sigma_\mu=2\sigma$. Remembering our previous discussion, we can now extract our first differential equation by requiring the coefficient of $\ln \si$ to vanish. Since this requirement has to hold independently of the differential operator chosen out of \eqref{Px}, we have \beq\label{firstH}P_xV=P_yV=D'_xD'_yV=(D'_x-D_y)V=(D'_y-D_x)V=0.\eeq To obtain further differential equations, we need to look at the terms involving $U$, {\it viz.}, 
$$P_x\left(\frac{U}{\sigma}\right)=\frac{\left(P_x
U\right)}{\sigma}+\frac{2\sigma_\mu U^\mu+(\square_x\sigma-4)U}{\sigma^
2},$$
which, combined with the $\sigma^{-1}$ coefficient coming from the series obtained out of differentiating
$V\ln\sigma$, leads us to the following two identities:
\begin{gather}
P_xU+2V_{0;\mu}\sigma^\mu+(\square_x\sigma-2)V_0=0,\label{secondH}\\
2U_{;\mu}\sigma^\mu+(\square_x\sigma-4)U=0,\label{thirdH}
\end{gather}
referring to the $\sigma^{-1}$ and $\sigma^{-2}$ coefficients, respectively.

Let us now focus on \eqref{thirdH}; one can infer that $U$ is subject to a linear partial differential equation which, according to
standard theorems, provides a unique solution once a suitable initial condition is given. The latter is
usually chosen in such a way that
$$[U]=I_4,$$ 
and, hence, the Cauchy problem associated to the $U$-bispinor strongly suggests us to hypothesise $U$ to be of the form $U=uI$, with a smooth biscalar $u$ satisfying $[u]=1$. Plugging in this ansatz in
\eqref{thirdH} and recalling the properties of $I$, it holds that $uI$ is the solution we are seeking if and only if $u$
satisfies the partial differential equation
$$2u_{;\mu}\si^\mu +(\square_x\sigma-4)u=0.$$ Hence, it turns out that $u$ fulfils the same
transport equation as the $\sigma^{-1}$ coefficients of the Hadamard bidistribution encoding the singularity of the two-point function for a scalar quantum field and is thus given by the square root of the so-called $Van Vleck-Morette$ determinant. It would be tempting to think that a similar result and interpretation holds for $V$, but, alas, this is far from being the truth as one can realise
by direct inspection of \eqref{secondH} since spin curvature terms not proportional to the identity enter the arena via derivatives of $I$. The enlarged complexity of the Diracian Hadamard coefficient $V$, however, is compensated by the increased number of differential equations fulfilled by $V$ \eqref{firstH}. Of course, any of them is enough to determine $V$, but computations are still considerably easier if one employs all.

 To obtain differential equations for the $V_n$, one has to combine
\eqref{Vexp2} with \eqref{firstH}. After a few formal manipulations, one gets to
$$\sum\limits_{n=0}^\infty\left(P_x V_n\right)\sigma^n+\sum\limits_{l=1}^\infty\left(2l V_{l;\mu}\sigma^\mu
+\left(l\square\sigma+2l\left(l-1\right)\right)V_l\right)
\sigma^{l-1}=0,$$ 
and, if we require this identity to hold true at each order in $\sigma$, to
\begin{gather}\label{fourthH}
P_x V_0+2V_{1;\mu}\sigma^\mu+(\square_x\sigma)V_1=0,\\
P_x V_n+2(n+1)V_{n+1;\mu}\sigma^\mu+\left(\left(n+1
\right)\square_x\sigma+2n\left(n+1\right)\right)(V_{n+1})=0.\quad\forall n\geq 1\label{fifthH}
\end{gather}
At this point it is clear how to determine $U$, $V$ and $W$ explicitly: the starting point is \eqref{thirdH}, which, as we have explained above, gives us $U$ once an initial condition has been assigned. Afterwards one can plug the result in
\eqref{secondH} in order to obtain $V_0$, though one needs to specify an initial condition.
This is already included in \eqref{secondH}, however, since, if we take the coincidence
point limit of \eqref{secondH} and recall the properties of $\si$, we end up with
$$[V_0]=-\frac{1}{2}[P_xU].$$
Hence, we can now proceed iteratively, namely, we exploit \eqref{fourthH} to construct $V_1$ once
we have specified the initial condition taking the coincidence point limit, {\it i.e.},
$$[V_1]=-\frac{1}{4}[P_xV_0].$$
Similarly, \eqref{fifthH} grants us that the same procedure allows us to express $V_{n+1}$ out
of the preceding term $V_n$ together with the initial condition
$$[V_{n+1}]=-\frac{1}{2(n+1)(n+2)}[P_xV_n].$$ Let us remember that all these results can be obtained starting from $[U]=I_4$ and employing differential equations which only involve local curvature terms and the mass $m$. Thus, both $U$ and $V$ indeed only depend on these data and are independent from the state under consideration. This of course changes once we want to determine the final unknown quantity $W$. Starting from \eqref{Wexp2} and recalling the differential equation $$P_x(H+W)=0\Leftrightarrow P_xW=-P_xH,$$ it is clear how to get recursive differential equations for the $W_n$. This time two initial conditions, namely, $[W_0]$ and $[W_{0;\mu}]$, however, have to be specified by hand, which is of course not surprising since we expect some indeterminateness which has to be fixed by selecting a specific state.

It seems that we finally have all ingredients necessary to calculate the sought coinciding point limits used in theorem \ref{tethe}. There is one potential feature of the Hadamard coefficients, however, which helps a lot simplifying calculations and should therefore be discussed before starting calculations, namely, their symmetry. Indeed, such a property has been proven in \cite{Moretti} for the scalar case, but, unfortunately, a similar result does not exist for Dirac fields and even understanding the correct notion of ``symmetry'' in our framework is a rather challenging task. We shall leave the tantalising endeavour to prove the symmetry of the Diracian Hadamard coefficients for possible future work and circumvent, for the time being, this gap with more explicit calculations. The following lemma will turn out to be rather useful in general and in the context of coping with the lack of (proven) symmetry in particular:

\lemma{\label{delho} Given a smooth bitensor $B(x,y)$ and a smooth biscalar $f(x,y)$  such that
$\frac{B(x,y)}{f(x,y)}$ is a smooth bitensor and $$[B]=[B_{;\mu'}]=[f]=[f_{;\mu'}]=0,\quad\text{as well as}\quad[
\square_y f]\neq 0,$$it holds $$\left[\frac{B}{f}\right]=\frac{[\square B]}{[\square f]}.$$}

\proof{We only sketch the proof here. Since $B$, $f$ and $B/f$ are smooth, their coinciding point limits do
not depend neither on $y$ nor on the path along which one approaches $x$. Thus, we can apply de l'Hospital's rule to our
smooth bitensors restricted to arbitrary smooth curves $\ga_i$, thereby expressing coinciding point limits
of fractions as those of directional derivatives, {\it e.g.},
$$\left[\frac{B(x,y)}{f(x,y)}\right]=\left[\frac{\dot{\ga_1}(y)^{\mu'} B(x,y)_{;\mu'}}{\dot{\ga_1}(y)^{\mu'} f(x,y)
_{;\mu'}}\right]=\left[\frac{\dot{\ga_1}(y)^{\mu'} \dot{\ga_2}(y)^{\nu'} B(x,y)_{;\mu'\nu'}}{\dot{\ga_1}(y)^{\mu'}\dot{
\ga_2}(y)^{\nu'} f(x,y)_{;\mu'\nu'}}\right]=\frac{[\dot{\ga_1}(y)^{\mu'} \dot{\ga_2}(y)^{\nu'} B(x,y)_{;\mu'\nu'}]}{[\dot{
\ga_1}(y)^{\mu'}\dot{\ga_2}(y)^{\nu'} f(x,y)_{;\mu'\nu'}]},$$ 
where we assume that $[\dot{\ga_1}(y)^{\mu'}\dot{\ga_2}(y)^{\nu'} f(x,y)_{;\mu'\nu'}]$ is non-vanishing. This holds due to the hypotheses of the lemma once we find two smooth curves $\ga_1$, $\ga_2$ such that $\dot{\ga_1}^{\mu'}\dot{\ga_2}^{\nu'}=g^{\mu'\nu'}$. Going to normal coordinates at $x$, it is always possible to find $\ga_1$, $\ga_2$ joining $x$ and $y$ such that $\dot{\ga_1}(x)
=(1,1,1,1)$ and $\dot{\ga_2}(x)=(-1,1,1,1)$; thus $\dot{\ga_1}^\mu\dot{\ga_2}^\nu=\eta^{\mu\nu}$, where
$\eta$ denotes the metric in normal coordinates, $\eta = \text{diag}(-1,1,1,1)$. Since $[\dot{\ga_1}(y)^{\mu'}\dot{\ga_2}(y)^{\nu'} f(x,y)_{;\mu'\nu'}]$ is coordinate-independent, the statement of the lemma holds as a consequence due to Synge's rule.}

The last worthy of mention tool to perform the calculations whose results we will display shortly is the computer. It should be clear at this point that there are lot of recursion relations to solve to achieve the wished-for results. Thus, at the least as a means of backing up manual calculations, computer algebra systems are a valuable instrument. To this avail, we have chosen to work with Mathematica and the free package \cite{Ricci}, suitable for performing calculations with vector bundles. The codes we have used to implement the recursive procedures and coinciding point limits are available upon request from \verb1t.p.hack@gmx.de1.
 
We can now finally state the main proposition of the appendix:

 \proposizione{\label{cplimits}The Hadamard bidistribution $H$ fulfils
 \begin{enumerate}
 \item $[P_xH]=6[V_1],\qquad[(P_xH)_{;\mu}]=8[V_{1;\mu}],\qquad [(P_xH)_{;\mu'}]=-8[V_{1;\mu}]+6[V_1]_{;\mu},$
 \item $[P_yH]=6[V_1],\qquad[(P_yH)_{;\mu}]=8[V_{1;\mu}]-2[V_1]_{;\mu},\qquad [(P_yH)_{;\mu'}]=-8[V_{1;\mu}]+8[V_1]_{;\mu},$
 \item $Tr[D'_xD'_yH]=-Tr[P_xH],\quad Tr[(D'_xD'_yH)_{;\mu}]=-Tr[(P_xH)_{;\mu}]+[V_1]_{;\mu},$ \\[5pt]
 $Tr[(D'_xD'_yH)_{;\mu'}]=-Tr[(P_xH)_{;\mu'}]-[V_1]_{;\mu},$
 \item $Tr[(P_yH-P_xH)_{;\mu'}]\ga^\mu\ga_\nu=2Tr[V_1]_{;\nu}.$
\end{enumerate}}

 \proof{\begin{enumerate}
\item We shall employ (\ref{firstH}), (\ref{secondH}),  and \eqref{thirdH}. These data entail
\beq\label{PxHsmooth}
P_xH = 2V_{1;\ro}\sigma^\ro+V_1(\square_x\sigma+2)+\cO(\si),
\eeq
and thus, taking the coinciding point limit and remembering those of $\si$ computed in
the previous section, $[P_xH]=6[V_1]$. Similarly, one gets, deriving (\ref{PxHsmooth}) once and performing
the limit, $[(P_xH)_{;\mu}]=8[V_{1;\mu}]$. By means of Synge's rule we finally have $[(P_xH)_{;\mu'}]=-8[
V_{1;\mu}]+6[V_1]_{;\mu}$.
\item
We would of course like to compute $P_yH$, but without any knowledge on the symmetries of the Diracian Hadamard coefficients, we have to verify the transport equations for
$P_y$, which otherwise would follow automatically from those for $P_x$ as it happens in the scalar case for the scalar Hadamard coefficients $u$ and $v$ \cite{Moretti}. To wit,
$$2U_{;\mu'}\sigma^{\mu'}+U(\square_y\sigma-4)=I\left(2u_{;\mu'}\sigma^{\mu'}+u(\square_y\sigma-4)\right)
=0,$$
where the first equality holds since the derivative of $I$ vanishes along the geodesic connecting $x$ and $y$ and the second one holds since $u(x,y)=u(y,x)$\footnote{The symmetry of $u$ does not have to be proved in the same long way as that of $v$ (see \cite{Moretti}), but it follows automatically by its explicit form $u(x,y)=\sqrt{det(\si_{\mu\nu'}(x,y))\sqrt{\abs{g(x)}^{-1}}\sqrt{\abs{g(y)}^{-1}}}$.} and $u$ is thus subject to transport equations for both $P_x$ and $P_y$. Since $P_yH$ is smooth, we now know that $$Z_1\doteq\frac{Y_1}{\si}\doteq\frac{P_yU+2V_{0;\mu}\sigma^\mu+V_0(\square_y\sigma-2)}{\si},$$ 
must be smooth too. Alas, it does not factorise into a term only involving the scalar coefficients $u$ and $v$ times $I$ and, up to now, we are unaware of a way to prove that it is identically vanishing. But we can try to compute whether it vanishes up to the derivative order we need for our purposes. To this end, it helps to split $V$ into $vI + \tilde{V}$, where $\tilde{V}$ is the non-trivial matrix part of $V$ stemming from the spin curvature. This way one can separate from $Y_1$ a term which vanishes due to the transport equation for $v$ and has to cope with the remainder only. Involved calculations yield 
$$[Y_1]=[Y_{1;\mu}]=[\square Y_1]=[(\square Y_1)_{;\mu}]=0,$$
and thus, employing lemma \ref{delho}, $[Z_1]=[Z_{1;\mu}]=0$.
Consequently, $$P_yH = 2V_{1;\ro'}\sigma^{\ro'}+V_1(\square_y\sigma+2)+\text{terms vanishing in the limit}$$ and $$(P_yH)_{;\mu} = 2V_{1;\ro'}\sigma^{\ro'}_{;\mu}+V_{1;\mu}(\square_y\sigma+2)+\text{terms vanishing in the limit}.$$ 
One can now straightforwardly obtain $[P_yH]=6[V_1]$, $[(P_yH)_{;\mu}]=8[V_{1;\mu}]-2[V_1]_{;\mu}$, and $[(P_yH)_{;\mu'}]=8[V_{1;\mu}]-8[V_1]_{;\mu}$.
\item  Let us define 
$$Z_2 \doteq (D_x-D'_y)H=(D_y-D'_x)H.$$ 
By direct inspection, 
$$D'_xD'_yH=-P_xH-D'_xZ_2.$$ 
Again we know that $Z_2$ is smooth and, alas, neither this quantity nor $D'_xZ_2$ turns out to be vanishing. Luckily enough, we can still extract some useful results at the level of traced coinciding point limits, at an order of derivatives high enough for our purposes. One computes 
\beq 
\begin{aligned} Z_2&=-\frac{U(D_x-D'_y)\si}{\si^2}+\frac{(D_x-D'_y)U-V(D_x-D'_y)\si}{\si}+\ln(\sigma)(D_x-D'_y)V\\
&\doteq-\frac{U(D_x-D'_y)\si}{\si^2}+\frac{Y_2}{\si}+\ln(\sigma)(D_x-D'_y)V.
\end{aligned}
\eeq 
As already discussed, the last term vanishes identically and so does the first term on account of  $$U(D_x-D'_y)\si=u(I\ga^\mu\si_\mu+\ga^{\mu'}I\si_{\mu'})=u(I\ga^\mu\si_\mu+I\ga^{\mu}g_\mu^{\mu'}\si_{\mu'})=0.$$ 
This leaves us with $Z_2=Y_2/\si$. Involved computations, employing $(D_x-D'_y)V=P_xV=0$ to exchange higher derivative terms with terms of lower derivative order in the appearing commutators with $\ga$-matrices, yield 
$$[Y_2]=[Y_{2;\mu}]=[\square Y_2]=0,\qquad [(\square Y_2)_{;\mu}]=6\left[[V_1],\ga_\mu\right].$$
After a few rearrangements and out of lemma \ref{delho}, one gets 
$$[Z_2]=0, \quad [Z_{2;\mu}]=\left[[V_1],\ga_\mu\right].$$ 
Hence, $[Z_{2;\mu}]$ is traceless due to the antisymmetry of the commutator. By means of formula \eqref{6gamma}, one can show per direct inspection that $Tr[V_1]\ga_\mu\ga_\nu = Tr[V_1]g_{\mu\nu}$ which entails that even $D'_xZ_2$ is traceless and, thus, 
$$Tr[D'_xD'_yH]=-Tr[P_xH].$$
In order to compute $Tr[(D'_xD'_yH)_{;\mu}]$ and $Tr[(D'_xD'_yH)_{;\mu'}]$, let us consider that $D'_yZ_2=D'_xZ_2+P_xH-P_yH$. Employing this as well as the previous results and tricks we have discussed in this proof, one obtains the following chain of identities 
\beq
\begin{aligned}
Tr[(D'_xZ_2)_{;\mu}]&=-Tr\ga^\nu[Z_{2;\nu\mu}]=Tr\ga^\nu[Z_{2;\nu'\mu}]-Tr\ga^\nu[Z_{2;\mu}]\nu\\ &=-Tr[(D'_yZ_2)_{;\mu}]=-Tr[(D'_xZ_2)_{;\mu}]+Tr[(P_yH-P_xH)_{;\mu}]\\ &=\frac{1}{2}Tr[(P_yH-P_xH)_{;\mu}]=-Tr[V_1]_{;\mu}\\ &=-Tr[(D'_yZ_2)_{;\mu'}].
\end{aligned}
\eeq 
We can finally use this last calculation to obtain 
$$Tr[(D'_xD'_yH)_{;\mu}]=-Tr[(P_xH)_{;\mu}]+[V_1]_{;\mu}, \qquad Tr[(D'_xD'_yH)_{;\mu'}]=-Tr[(P_xH)_{;\mu'}]-[V_1]_{;\mu}.$$
\item  Inserting the previous results, we have $[(P_yH-P_xH)_{;\nu'}]=2[V_1]_{;\nu}.$ As already discussed, due
to $\ga$-matrix identities, tracing $[V_1]$ with two $\ga$-matrices amounts to a multiplication with the
metric. Since the operations of trace and covariant derivation commute, we have $Tr[V_1]_{;\nu}\ga^\nu\ga^\mu=Tr[V_1
]_{;\mu}$ and thus
$$Tr[(P_yH-P_xH)_{;\nu'}]\ga^\nu\ga^\mu=2[V_1]_{;\mu}.$$
\end{enumerate}}

We would like to conclude this section by stating the last ingredient necessary for proving theorem \ref{tethe}, the coinciding point limit of $V_1$, {\it viz.},
$$[V_1]=\left(\frac{m^4}{8}+\frac{m^2R}{48}+\frac{R^2}{1152}+\frac{\square R}{480}-\frac{R_{\alp\be}R^{\alp\be}}{720}+\frac{R_{\alp\be\ga\de}R^{\alp\be\ga\de}}{720}\right)I_4+\frac{C_{\alp\be}C^{\alp\be}}{48}.$$

\subsection{Conserved local curvature tensors}

The explicit form of the conserved local curvature tensors spanning the regularisation freedom of the expected stress-energy tensor is:

\beq\begin{aligned} I_{\mu\nu}& \doteq \frac{1}{\sqrt{|g|}}\frac{\delta}{\delta g_{\mu\nu}}\int\limits_M R^2 d\mu_g  = g_{\mu\nu}\left(\frac{1}{2}R^2+2\square R\right)-2R_{;\mu\nu}-2R R_{\mu\nu},\notag\\
J_{\mu\nu}&\doteq \frac{1}{\sqrt{|g|}}\frac{\delta}{\delta g_{\mu\nu}}\int\limits_M R_{\alp\be} R^{\alp\be} d\mu_g= \frac{1}{2}g_{\mu\nu}(R_{\mu\nu} R^{\mu\nu}+\square R)-R_{;\mu\nu}+\square R_{\mu\nu} -2 R_{\alp\be}R^{\alp\,\,\,\be}_{\,\,\,\mu\,\,\,\nu},\notag\\
 K_{\mu\nu}&\doteq \frac{1}{\sqrt{|g|}}\frac{\delta}{\delta g_{\mu\nu}}\int\limits_M R_{\alp\be\ga\de} R^{\alp\be\ga\de} d\mu_g\notag\\ &= -\frac{1}{2}g_{\mu\nu}R_{\alp\be\ga\de} R^{\alp\be\ga\de}+2R_{\alp\be\ga\mu} R^{\alp\be\ga}_{\,\,\,\,\,\,\,\,\,\nu}+4 R_{\alp\be}R^{\alp\,\,\,\be}_{\,\,\,\mu\,\,\,\nu}-4R_{\alp\mu}R^{\alp}_{\,\,\,\nu}-4\square R_{\mu\nu}+2R_{;\mu\nu}.\notag\end{aligned}\eeq

As already stated, in four spacetime dimensions, these are related as $K_{\mu\nu}=I_{\mu\nu}-4 J_{\mu\nu}$ via the generalised Gauss-Bonnet-Chern theorem \cite{alty, thooft}. Furthermore, in this case, they all have a trace proportional to $\square R$ and, thus, the linear combination $I_{\mu\nu}-3 J_{\mu\nu}$ is traceless.

\end{document}